\documentclass{aa}
\usepackage{newtxtext,newtxmath}
\usepackage[T1]{fontenc}
\usepackage{ae,aecompl}
\usepackage{xcolor}
\bibpunct{(}{)}{;}{a}{}{,}
\usepackage[colorlinks=true, citecolor=blue]{hyperref}
\usepackage{multirow}
\makeatletter
\renewcommand*\aa@pageof{, page \thepage{} of \pageref*{LastPage}}
\makeatother
\usepackage{etoolbox}
\usepackage{graphicx}	
\usepackage{amsmath}	
\usepackage{lastpage}
\usepackage{verbatim}
\usepackage{units}
\usepackage{pdflscape}
\usepackage{svg}
\usepackage{bm}
\usepackage{stfloats}
\usepackage{float}
\usepackage{multicol}
\usepackage{hyperref}
\usepackage{multirow}
\usepackage{wasysym}
\usepackage{subcaption}
\usepackage{dcolumn}
\newcolumntype{d}[1]{D{.}{.}{#1}}
\newcolumntype{p}[1]{D{,}{\pm}{#1}}
\newcommand\mc[1]{\multicolumn{1}{c}{#1}}

\newcommand{\degree}{\ensuremath{^\circ}}
\newcommand{\arcsecfrac}{\(\stackrel{\:''}{\textstyle.}\)}

\newcommand{\msun}{${\mathrm{M}}_{\odot}$}

\begin{document}

\title{Strongly polarised radio pulses from a new white-dwarf-hosting long-period transient}

\author{S.~Bloot$^{1,2}$\thanks{Corresponding author \email{bloot@astron.nl}}, H.~K.~Vedantham$^{1,2}$, C.~G.~Bassa$^{1}$, J.~R.~Callingham$^{1,3}$, W.~M.~J.~Best$^{4}$, M.~C.~Liu$^{5}$, E.~A.~Magnier$^{4}$, T.~W.~Shimwell$^{1,6}$, T.~J.~Dupuy$^{7}$}
\authorrunning{S.~Bloot et al.}
\institute{$^{1}$ASTRON, Netherlands Institute for Radio Astronomy, Oude Hoogeveensedijk 4, Dwingeloo, 7991\,PD, The Netherlands\\
$^{2}$Kapteyn Astronomical Institute, University of Groningen, P.O. Box 800, 9700 AV, Groningen, The Netherlands\\
$^{3}$Anton Pannekoek Institute for Astronomy, University of Amsterdam, 1098 XH, Amsterdam, the Netherlands\\
$^4$The University of Texas at Austin, Department of Astronomy, 2515 Speedway, C1400, Austin, TX 78712, USA\\
$^5$Institute for Astronomy, University of Hawaii, 2680 Woodlawn Drive, Honolulu, HI 96822, USA\\
$^{6}$Leiden Observatory, Leiden University, PO Box 9513, Leiden, 2300 RA,
The Netherlands\\
$^{7}$Institute for Astronomy, University of Edinburgh, Royal Observatory, Blackford Hill, Edinburgh, EH9 3HJ, UK\\
}
\date{Received XXX; accepted YYY}

\label{firstpage}

\abstract{Long-period transients (LPTs) are a new and enigmatic class of objects that produce bright pulsations in the radio, with periods far exceeding those seen in rotationally powered pulsars. The proposed progenitors for LPTs are contested, with white dwarfs or magnetars being likely candidates. Here, we present the discovery of ILT\,J163430+445010, a new LPT detected in a blind search for Stokes\,V transients in the LOFAR Two-Metre Sky Survey. 
Unusual for LPTs, J1634+44 shows pulses that are 100\% circularly polarised, as well as pulses that are 100\% linearly polarised, with the polarisation state changing from pulse to pulse. We detect 19 pulses in total, each with a total polarisation fraction of $\sim100\%$ and a pulse duration of at most 10\,s. The pulses show a periodicity at $841.24808\pm0.00015$\,s, implying a low duty cycle of $0.012$.
J1634+44 has a marginally detected counterpart in the ultraviolet GALEX MIS survey and the ultraviolet/optical UNIONS survey, suggesting that it contains a white dwarf with an effective temperature between 15000\,K and 33000\,K. We do not detect J1634+44 with a deep $J$-band exposure with UKIRT at a $3\sigma$ AB magnitude limit of 24.7, ruling out a main-sequence star or ultracool dwarf with a spectral type earlier than M7.
The pulses from J1634+44 follow a particular pattern, with two pulses being produced every five periods after a waiting time of two or three periods. This pattern could be a result of spin-orbit coupling in a binary system with a 5:2 or 5:3 resonance, where a companion induces beamed radio emission on the white dwarf. The companion is most likely an ultracool dwarf or another white dwarf, making J1634+44 unique among the currently known sample of LPTs.
}

\maketitle
\nolinenumbers
\section{Introduction}
\noindent Over the last few years, an enigmatic class of radio sources has emerged that are referred to as long-period transients (LPTs) \citep[e.g.][]{2005Natur.434...50H,2022Natur.601..526H}. These LPTs emit short, bright pulses at periods ranging from a few minutes to a few hours \citep[e.g.][]{2024arXiv240707480D, 2025NatAs.tmp...36L}. At the time of writing, only nine LPTs have been found \citep{2005Natur.434...50H,2022Natur.601..526H,2023Natur.619..487H, 2024NatAs...8.1159C,2024arXiv240707480D,iris,hurley-walker24, 2024arXiv241116606W,2024arXiv241115739L, 2025NatAs.tmp...36L}. Based on distance estimates from their cold plasma dispersion, these objects must be located in the Milky Way. Their high radio brightness temperatures and polarised fractions imply that the pulses are produced through a coherent plasma process. However, the exact emission process that is producing the radio emission is yet to be identified. 

The origins of LPTs are hypothesised to be compact objects, either extremely slowly rotating magnetars \citep[e.g.][]{beyond_deathline}, or white dwarfs \citep[e.g.][]{2005ApJ...631L.143Z, katz_2022}. Magnetars are known to produce radio emission at longer periods than canonical pulsars, up to a few seconds \citep[see e.g.][for a review]{2017ARA&A..55..261K}. Since the emission is powered by their strong magnetic field ($B\sim 10^{15}\,{\rm G}$) instead of rotation, it is possible for even slower magnetars, with periods approaching hours, to still produce bright radio pulses \citep{beyond_deathline}. While such objects have been hypothesised as a potential origin for LPTs, there has been no conclusive evidence. Most known magnetars produce stochastic radio emission accompanying an X-ray outburst. However, only one of the LPTs followed up at X-ray wavelengths has shown any emission \citep{2024arXiv241116606W}, despite extensive searches \citep{2022Natur.601..526H, 2023Natur.619..487H, 2024NatAs...8.1159C, 2024arXiv240707480D,iris, Rea_2022}.

A different origin for LPTs was proposed by \citet{2005ApJ...631L.143Z}, who suggested that LPTs may be isolated white dwarfs. White dwarfs are not previously known to produce rotationally-powered radio pulses, although it may be theoretically plausible to generate such emission through a mechanism similar to that operating in canonical pulsars \citep[e.g.][]{2005ApJ...631L.143Z}. 

Most LPTs have thus far been detected in the Galactic plane, which have made their nature difficult to determine due to high levels of extinction and crowding preventing the identification of optical counterparts. However, recently two LPTs were found at high Galactic latitudes that allowed for the convincing identification of their optical counterparts \citep{iris, hurley-walker24}. In both cases, the optical counterpart was a main-sequence star. \citet{iris} found a blue excess in the colours of the M\,dwarf counterpart and large radial velocity shifts that led them to conclude that the LPT is a white dwarf in a binary with a main-sequence star. The optical counterpart to GLEAM-X J0704‑37, another LPT at high Galactic latitude found by \citet{hurley-walker24}, also appears to be a main-sequence star. Spectroscopic follow-up by \citet{2025arXiv250103315R} has shown that the observed radial velocities of GLEAM-X J0704‑37 are also consistent with a binary composed of an M\,dwarf and a white dwarf.

Here, we present a new LPT, ILT\,J163430+445010 (J1634+44 hereafter), discovered in the LOFAR Two Metre Sky Survey \citep[LoTSS][]{lotss-dr2} in an untargeted search for circularly polarised pulses (Bloot et al., in prep.). J1634+44 has a period of $\sim$14\,min, is located far off the Galactic plane at a Galactic latitude of 40\degree, and, unusually, emits pulses that are $\sim$100\% polarised. Its properties are summarised in Table\,\ref{tab:properties}. We discuss how this new source fits within the emerging paradigm of LPTs. 

J1634+44 was also independently discovered with the Canada Hydrogen Intensity Mapping Experiment (CHIME). Their data and interpretation are presented in \citet{adam}.

\section{Radio observations}
\label{sec:radio}
We first discovered J1634+44 in a search for Stokes\,V transients in the second data release of the LOFAR Two-Meter Sky Survey \citep[LoTSS DR2,][]{2022A&A...659A...1S}. LoTSS DR2 consists of 8-hour exposures covering 27\% of the Northern sky at a frequency of 120-167\,MHz. We searched this dataset in Stokes\,V image space by making snapshot images with integration times of 60, 20, and 4 minutes (full survey description in Bloot et al. (in prep.)). Previous searches of the LoTSS DR2 data have only searched for Stokes\,V sources in full 8\,h integrations \citep{vlotss}.

\begin{figure}
    \centering
    \includegraphics[width=0.95\linewidth]{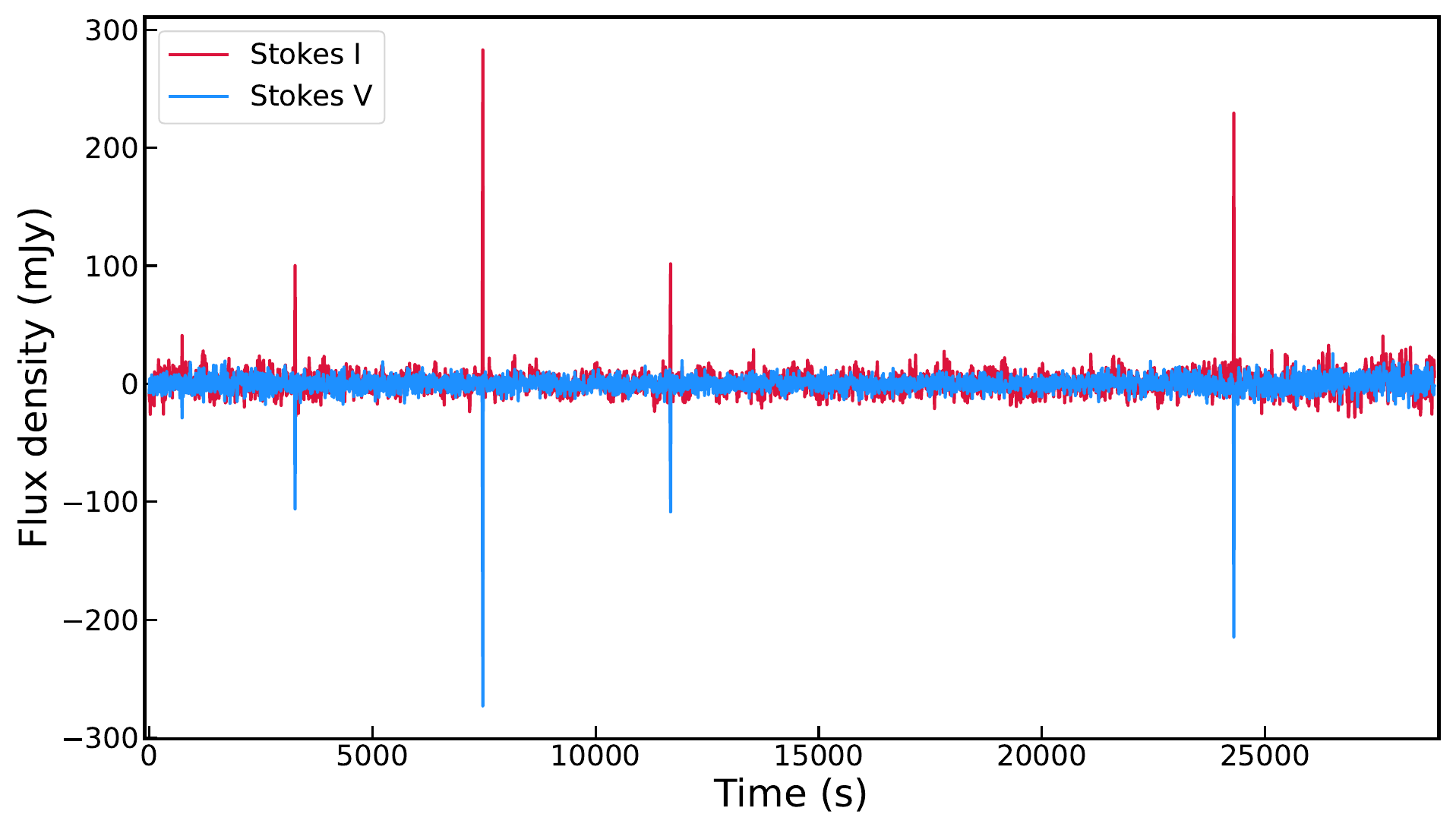}
    \caption{Light curve with 8\,s time resolution of the LOFAR observation where J1634+44 was first detected, in Stokes I (red) and Stokes V (blue). The time axis shows the time since the start of the observation in seconds.}
    \label{fig:lc}
\end{figure}

J1634+44 was detected in the Stokes\,V transient search as a 8.7\,$\sigma$ source, corresponding to 12.5$\pm$1.4\,mJy, in a single 4\,minute snapshot in Stokes\,V, in LoTSS pointing \texttt{P247+43}, $\approx$2.3 degrees from the pointing centre. The significance increased to 18$\sigma$ when we imaged the data set at a cadence of 1 minute, indicating that the emission had a small temporal duty ratio. The images were made using \texttt{WSclean} \citep{2014MNRAS.444..606O}, with a robust parameter of 0.0 for Stokes\,V and -0.5 for Stokes\,I. Only baselines with lengths between 200\,m and 90\,km were included.
We subtracted the other sources in the field using the extraction procedure described in \citet{ 2021A&A...651A.115V}. We then created a light curve of the full observation at the location of J1634+44 in both Stokes\,V and Stokes\,I (Fig.\,\ref{fig:lc}), following the method described in \citet{bloot2024}.
These light curves at 8\,s resolution revealed five short ($<$16\,s) pulses over the 8\,hr observing window. J1634+44 lies within the main lobe of the primary beam in 8 LoTSS pointings, amounting to six independent observations, as two observations consisted of two simultaneously observed pointings. We detected a total of 19 pulses spread over five of these observations, with no detections in the sixth observation.
We list the times of arrival (TOAs), the flux densities, and the polarisation fractions of the pulses in Table\,\ref{tab:toas}. We use the Stokes\,V sign convention used in pulsar astronomy, where Stokes\,V is defined as left-handed circularly polarised light minus right-hand circularly polarised light. 
As is clear from the table, the flux density of J1634+44 is strongly variable. All pulses are strongly polarised, with some being strongly circularly polarised and others strongly linearly polarised.

\begin{table}[]
    \def\arraystretch{1.25}
    \centering 
    \caption{Measured and inferred properties of J1634+44.}
    \resizebox{\linewidth}{!}{%
    \begin{tabular}{lc}
    \hline \hline
    Parameter & Value \\
    \hline \hline
         Right ascension & 16h34m30s $\pm$2" \\
         Declination & +44\degree50'10" $\pm$2"\\
         Galactic longitude ($l$) & 70.1692\degree $\pm$2"\\
         Galactic latitude ($b$) & 42.5754\degree $\pm$2" \\
         Maximum flux density & 383$\pm$52\,mJy \\
         Pulse period & 841.24808$\pm$0.00-15\,s\\
         Period derivative & $-9\pm3 \times10^{-12}$\\ 
         Modulating period & 2103.110$\pm$0.002\,s \\
         Dispersion measure & 22.5$\pm$5.5 pc cm$^{-3}$\\
         Rotation measure & $6.27\pm$0.04\,rad\,m$^{-2}$ \\
         Distance & 1.0-4.3\,kpc\\
         Maximum radio luminosity at 1\,kpc & $(4.5\pm0.6)\times10^{20}$\,erg\,s$^{-1}$\\
         Maximum radio luminosity at 4.3\,kpc & $(85\pm1)\times10^{20}$\,erg\,s$^{-1}$\\
         \hline
    \end{tabular}
    }
    \tablefoot{The co-ordinates listed here are measured for the brightest pulse in Stokes\,V. The error given on the co-ordinates represents $3\sigma$ uncertainty on the localisation. The maximum flux density and radio luminosity are the Stokes\,V flux density of the brightest pulse, integrated over the full bandwidth of the observation. The distance is derived from the dispersion measure following the Galactic electron density models by \citet{2002astro.ph..7156C} and \citet{2017ApJ...835...29Y}.}
    \label{tab:properties}
\end{table}

\begin{table*}[]
    \def\arraystretch{1.25}
    \centering
    \caption{Time of arrival of the detected pulses.}
    \begin{tabular}{lcp{3.3}p{3.3}p{3.3}p{4.4}p{4.4}ll}
    \hline \hline
    \mc{Time (UTC)} & LoTSS pointing & \mc{S$_{I}$ (mJy)} & \mc{S$_{V}$ (mJy) } & \mc{S$_{L}$ (mJy)} & \mc{S$_{V}$/S$_{I}$} & \mc{S$_{L}$/S$_{I}$} & \mc{Rotation out of 5}\\
    \hline \hline
2015-09-02T17:05:02 & P248+48 & 120,25 & -134,17 & 73,12 & -1.12,0.28 & 0.61,0.16 & 4 \cr
\hline
2016-03-25T00:05:57 & P247+43 & 41,7 & -29,7 & 21,7 & -0.71,0.2 & 0.53,0.2 & 1 \cr
2016-03-25T00:48:09 & P247+43 & 100,7 & -106,6 & 35,6 & -1.06,0.09 & 0.35,0.07 & 4 \cr
2016-03-25T01:58:14* & P247+43 & 283,6 & -273,6 & 37,5 & -0.97,0.03 & 0.13,0.02 & 4 \cr
2016-03-25T03:08:20 & P247+43 & 102,5 & -109,5 & 16,4 & -1.07,0.07 & 0.15,0.04 & 4 \cr
2016-03-25T06:38:37 & P247+43 & 229,8 & -215,7 & 30,6 & -0.94,0.04 & 0.13,0.03 & 4 \cr
\hline
2016-05-18T21:34:19 & P243+45 & 179,36 & -234,23 & 85,28 & -1.31,0.29 & 0.47,0.18 & 4 \cr
2016-05-18T22:44:17 & P243+45 & - & -99,25 & - & - & - & 4 \cr
2016-05-18T23:54:23 & P243+45 & 151,33 & -130,21 & - & -0.86,0.24 & - & 4 \cr
2016-05-19T03:24:48 & P243+45 & 383,52 & -346,33 & - & -0.9,0.15 & - & 4 \cr
\hline \hline
2019-05-24T20:41:23 & P251+43 & 184,18 & - & 180,16 & - & 0.98,0.13 & 3 \cr
2019-05-24T21:23:35 & P251+43 & 163,18 & -93,12 & 49,15 & -0.57,0.1 & 0.3,0.09 & 1 \cr
2019-05-24T22:33:41 & P251+43 & 302,17 & -273,10 & 65,10 & -0.9,0.06 & 0.21,0.04 & 1 \cr
2019-05-24T23:01:35 & P251+43 & 129,14 & - & 122,14 & - & 0.94,0.15 & 3 \cr
2019-05-25T00:11:41 & P251+43 & 142,23 & - & 65,12 & - & 0.46,0.11 & 3 \cr
\hline
2019-06-06T20:39:21 & P251+45 & 152,5 & -53,4 & 135,4 & -0.35,0.03 & 0.89,0.04 & 3 \cr
2019-06-06T21:49:27 & P251+45 & 51,4 & -10,3 & 44,4 & -0.19,0.06 & 0.85,0.11 & 3 \cr
2019-06-07T02:01:56 & P251+45 & 131,6 & -140,4 & - & -1.06,0.06 & - & 1 \cr
2019-06-07T02:29:51 & P251+45 & 91,6 & -21,5 & 84,5 & -0.23,0.06 & 0.92,0.09 & 3 \cr
         \hline \hline
    \end{tabular}
    \tablefoot{Time of arrival of the pulses in seconds of modified Julian date, the LoTSS pointing in which it was observed, peak flux density in Stokes\,I (S$_{I}$), Stokes\,V (S$_{V}$), the absolute value of Stokes\,Q + i\,Stokes\,U (S$_{L}$) in mJy, and the circular and linear polarisation fraction for all pulses detected on J1634+44. For each pulse, we also list the rotation number out of five. The uncertainty on all TOAs is taken to be 4\,s, as the TOAs are based on data with a time resolution of 8\,s. The pulse marked with * is the pulse with the highest signal-to-noise ratio. A machine-readable version of this table is available at \url{https://github.com/SBloot/J1634-44}.}
    \label{tab:toas}
\end{table*}

\begin{figure}
    \centering
    \includegraphics[width=0.95\linewidth]{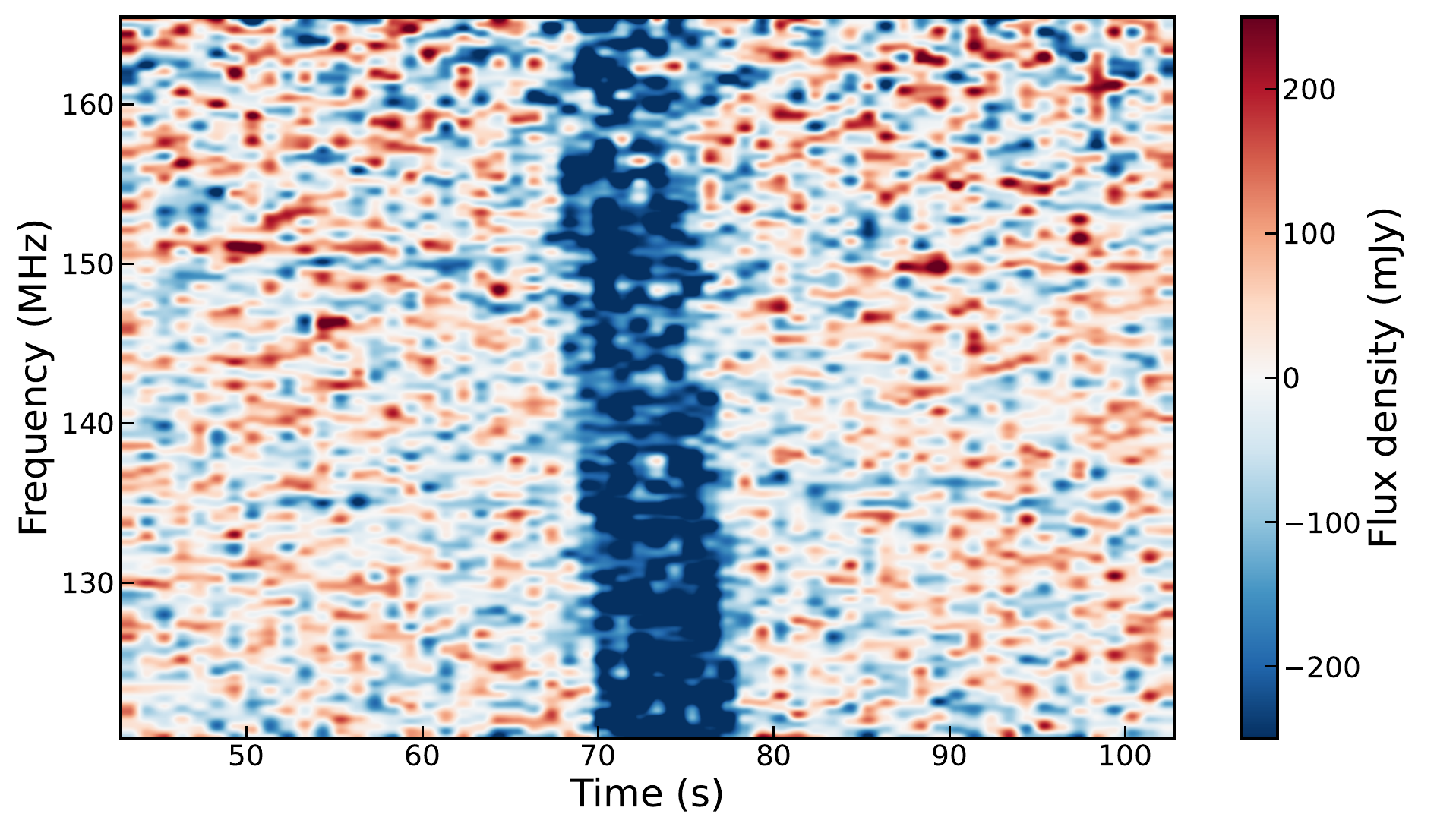}
    \caption{Stokes\,V dynamic spectrum of the highest signal-to-noise ratio pulse (marked with * in Table\,\ref{tab:toas}) with LOFAR on J1634+44. The dynamic spectrum is smoothed with a Gaussian filter along the frequency axis with a sigma of 244\,kHz. The time resolution is 1\,s. The time axis shows the time offset from 2016-03-25 01:57:00 UTC.}
    \label{fig:pulse}
\end{figure}

Due to the transient nature of the source and residual ionospheric effects, we were only able to localise the source with a 2" uncertainty in each epoch. The sets of co-ordinates agree within 2\,$\sigma$ across the 3.8\,yr timespan covered by the observations. We therefore set an upper limit on the proper motion of J1634+44 of 1"\,yr$^{-1}$.

\subsection{Dispersion measure}
\begin{figure}
    \centering
    \includegraphics[width=0.95\linewidth]{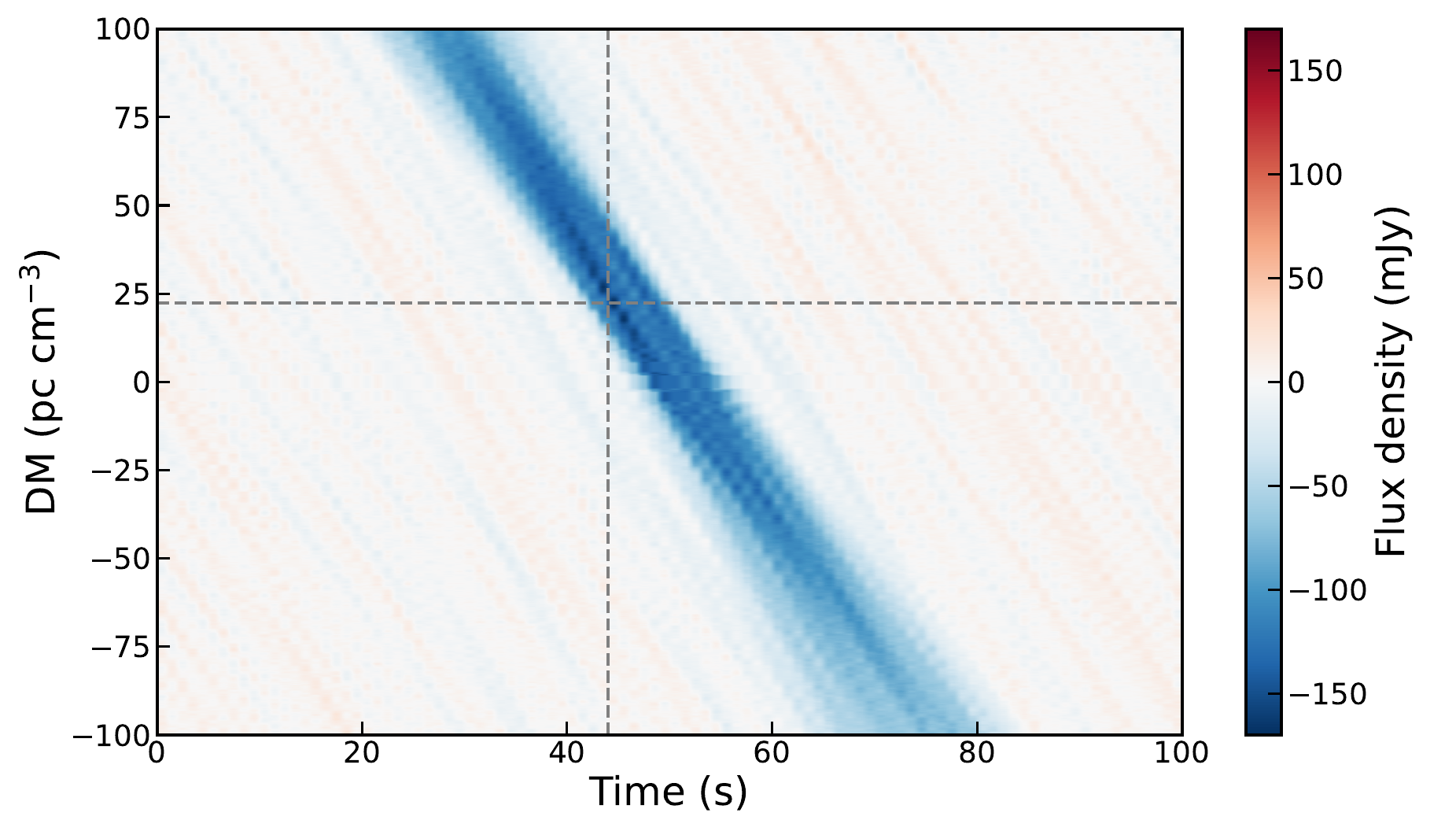}
    \caption{Bow-tie plot of the brightest pulse detected from J1634+44 with LOFAR in Stokes\,V. The crosshairs indicate the location of maximum signal-to-noise, at a DM of 22\,pc\,cm$^{-3}$.}
    \label{fig:bowtie}
\end{figure}

The LoTSS pointings were originally recorded with a time resolution of 1\,s and then averaged to 8\,s by the standard survey calibration. We reprocessed the pointing containing the brightest pulse, \texttt{P247+43}, using the \texttt{LINC} pipeline \citep{linc} without time averaging, to obtain a dynamic spectrum at 1\,s resolution. 
With the data reprocessed at 1\,s, we can see the structure of the pulses in detail. Each pulse lasts for no more than $10$\,s, leading to an extremely low duty cycle of $0.012$. The pulses show a time-frequency sweep, as shown in Fig.\,\ref{fig:pulse}, with a delay of $\sim$2\,s across the bandwidth of 47\,MHz. The sweeping structure agrees well with the expected shape of cold plasma dispersion. To find the dispersion measure (DM), we calculated the SNR of the brightest pulse after averaging across frequency for a range of DM trials. Averaging in frequency is possible because the sign of Stokes\,V is consistent across the pulse.
We find that the dispersion measure (DM) corresponding to the maximum SNR is 22\,pc\,cm$^{-3}$. The corresponding `bow-tie' plot is shown in Fig.\,\ref{fig:bowtie}.

To determine the uncertainty on the DM, we averaged two sections of 10\,MHz of clean bandwidth, centred at 126\,MHz and at 156\,MHz, and determined the delay between the start time of the pulse in both frequency bins. We defined the start time of the pulse as the first time bin that exceeds 5$\sigma$. We found that the delay is 2$\pm$0.5\,s, which corresponds to a DM of $22.5\pm5.5$\,pc\,cm$^{-3}$. 
Using the Galactic electron distribution models from both \citet{2002astro.ph..7156C} and \citet{2017ApJ...835...29Y}, this DM range corresponds to a distance range of 1.0--1.6\,kpc and 1.5--4.3\,kpc respectively, at the co-ordinates of J1634+44. We conclude that the distance to J1634+44 is in the range of 1--4.3\,kpc.

\subsection{Timing solution}
The pulses detected from J1634+44 are strongly periodic, as can already be seen from the single observation shown in Figure\,\ref{fig:lc}. To determine a timing solution, we use \texttt{PINT} \citep{pint} with several different timing models. We use the DE436 ephemeris and we include the Solar System Shapiro delay. The results of each model are shown in Table\,\ref{tab:timing-models}, and the timing residual plots are included in Appendix\,\ref{app:tr}.

First, we fit a simple model with only the period as a free parameter. We started by fitting this model to only the observation on MJD $58627-58628$, which resulted in a period of $840.8\pm0.4$\,s. When we extended the fit to include all data in 2019, we found a period of $841.246\pm0.002$\,s, with residuals up to 6\,s. Using this period and applying the solution to all pulses, we found residuals up to 10\,s. A fit with all time of arrivals (TOAs) included did not converge. When we included a period derivative ($\dot{P}$) in the model, the fit did converge, with a period of $841.24808\pm0.00015$\,s and a $\dot{P}$ of $-9\pm3\times10^{-12}$\,s\,s$^{-1}$.

From the light curves of individual epochs, we see that the pulses seem to arrive roughly every $70$\,minutes, with some pulses offset compared to that periodicity by about $28$\,minutes. Such an arrangement is reminiscent of pulsar TOAs with a main pulse and interpulse. We therefore fit a model with two free parameters, the period and a phase offset between a main pulse and an interpulse, to the MJD $58627-58628$ data, resulting in a period of 4205.8$\pm$3.3\,s and a phase jump of $0.601\pm0.0015$. When we extended the fit to all 2019 data, we found a best-fit period of $4206.24\pm0.01$\,s, with an offset for the interpulse of $0.601\pm0.001$. However, when we extended the model to the full dataset, it did not provide a good fit, with residuals of up to 500\,s. When we plotted the residuals using the best solution on only the 2019 data, we saw that the pulses in 2015--2016 seem to be offset compared to the pulses in 2019 by $0.6$ in phase. To account for this offset, we added another jump to the model, between the 2015--2016 and the 2019 pulses. The best-fit solution has a period of 4206.2406$\pm$0.0008\,s, a phase offset between the main pulse and the interpulse of $0.6015\pm0.0005$ and a jump between the 2015--2016 and 2019 pulses of $0.614\pm0.005$.

\begin{table*}[]
\caption{Timing solutions for J1634+44.}
\resizebox{\linewidth}{!}{%
    \def\arraystretch{1.25}
    \centering 
    \begin{tabular}{lrp{9.9}p{7.7}p{5.1}p{5.2}p{2.2}}
    \hline \hline
    Model & $\chi^2_{\nu}$ & \mc{Period (s)} & \mc{Interpulse phase} & \mc{Phase jump} & \mc{$\dot{P}$ ($10^{-12}$ s\,s$^{-1}$)} \\
    \hline \hline
    Short period & 1.02 &  841.246,0.002 & - & -  & - \cr
    Long period with interpulse & 0.62 & 4206.24,0.01 & 0.601,0.001  & -  & -\cr 
    Short period with $\dot{P}$ & 0.95 & 841.24808,0.00015 & - & - & -9,3 \cr
    Long period with interpulse and phase offset & 0.91 & 4206.2406,0.0008 & 0.6015,0.0005 & 0.614,0.005  & -\cr
         \hline
    \end{tabular}%
    }
    \tablefoot{Results of the various timing models described in the text. Note that the reduced chi-squared value ($\chi^2_{\nu}$) is calculated using only the data in 2019 for the first two timing models, and using all data in the last two models.}
    \label{tab:timing-models}
\end{table*}

\subsection{Waiting time analysis}
\begin{figure}
    \centering
    \includegraphics[width=0.95\linewidth]{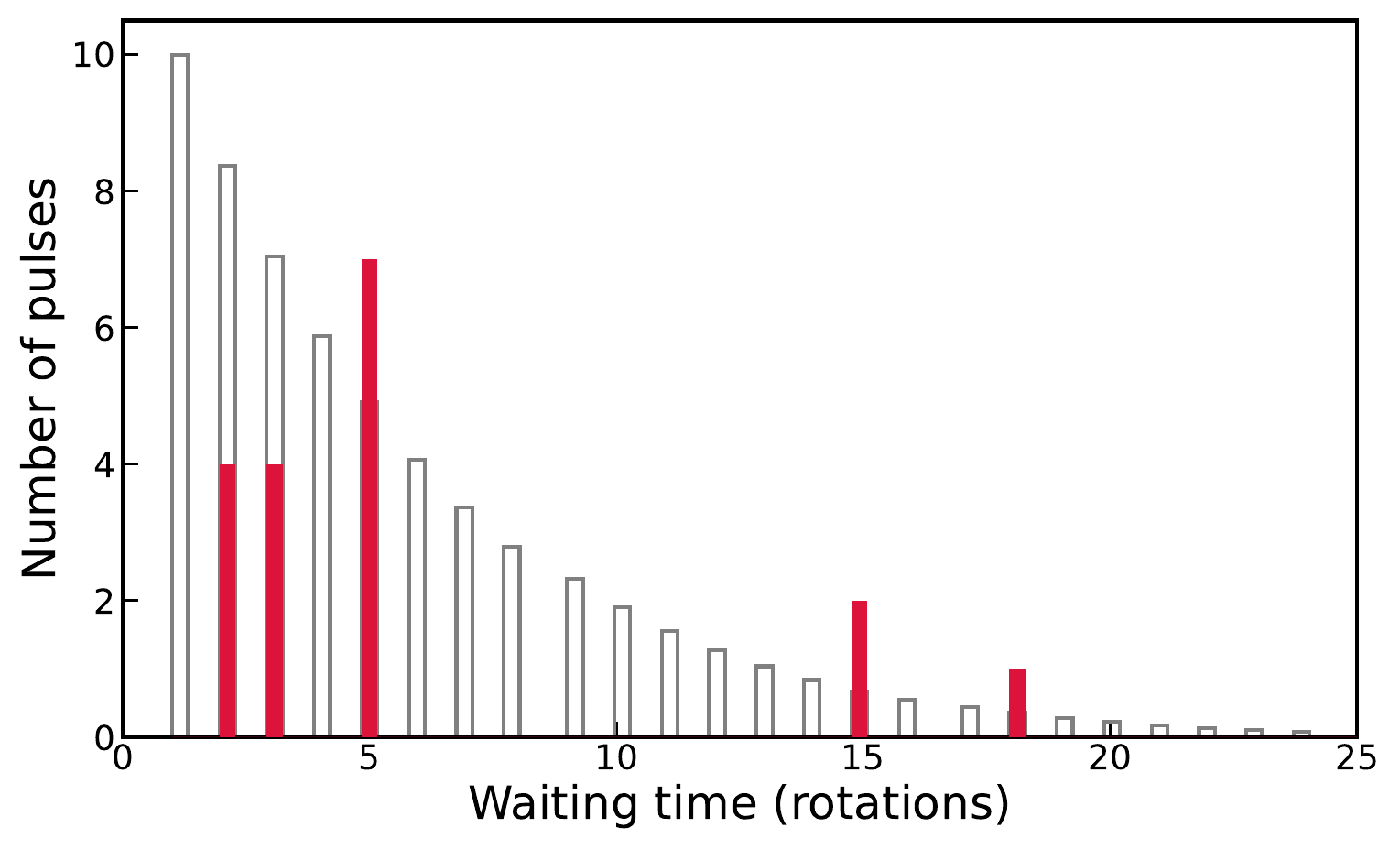}
    \caption{Distribution of waiting times between pulses for trials assuming the pulses are randomly distributed across the observing time (grey) and for the observed TOAs (red), using a period of 841\,s.}
    \label{fig:wait-times}
\end{figure}

\begin{figure*}
    \centering
    \includegraphics[width=0.95\linewidth]{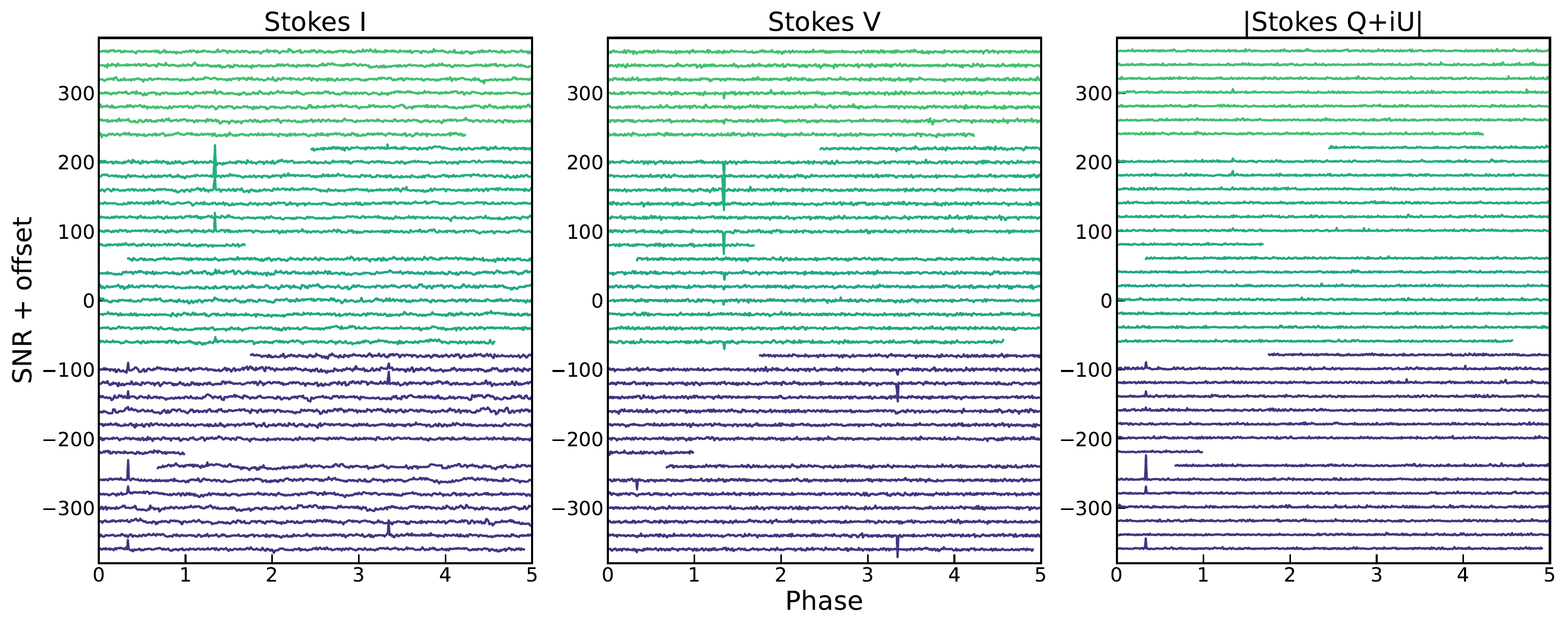}
    \caption{Light curve of all observations, binned to 8 seconds, phase-wrapped using a timing solution with a period of 841\,s and a period derivative. The phase zero point is arbitrarily set to the start of the first observation. The left panel shows Stokes\,I, the middle panel shows Stokes\,V, and the right panel shows the linear polarisation. The data is plotted chronologically, with the earliest observation at the top and the latest observation at the bottom of the plot. The lines are coloured according to the time of observing, starting with green for the first observation.}
    \label{fig:lc_wrapped}
\end{figure*}

The best-fit timing solutions contain a period of 841 seconds, or a period of 4206 seconds with a phase offset of 0.6 between a main pulse and an interpulse, and a jump of 0.6 in phase between 2015--2016 and 2019. The jump values and the $4206$\,s period are all suspiciously close to integer multiples of the $841$\,s period suggesting that the 841\,s period is the `true' period and the jumps originate from an additional `on-off' periodic modulation of the pulses. To test this hypothesis, we performed a waiting time analysis. For each pulse, we determined the number of periods it takes for the next pulse to arrive. The distribution of waiting times using the 841\,s period is plotted in Figure\,\ref{fig:wait-times}. We have zero waiting times of one period, several at two and three periods, none at four periods, and seven at five periods. Five periods corresponds exactly to the longer period of 4206 seconds, and two and three rotations correspond to the phase offset of 0.4. If we randomly distribute 20 pulses across four 8-hr observations, we can produce a distribution of waiting times with no waiting times of one period in only 5\% of trials. A distribution with no waiting times of one and four periods only occurs in 0.5\% of trials. We therefore conclude that the pulses do not arrive randomly, but instead at specific integer numbers of the 841\,s period. The timing solution with the 4206\,s period and the two phase offsets describes the data well because the phase offsets are integer multiples of the shorter 841\,s period. We plot the light curves of all observations phase-wrapped to the 841\,s period in Fig.\,\ref{fig:lc_wrapped}.

\subsection{Polarisation}
The polarisation properties of J1634+44 vary from pulse to pulse. Some pulses are 100\% circularly polarised, whereas others are strongly linearly polarised. We select the three pulses with the strongest linear polarisation and determine the rotation measure (RM) for each of them, using \texttt{rm-tools} \citep{2020ascl.soft05003P}. We take the mean of these RMs to find an average RM for J1634+44 of $6.27\pm0.04$\,rad\,m$^{-2}$.
The linear polarisation in Table\,\ref{tab:toas} was determined by de-rotating the dynamic spectrum using this value, averaging across frequency, and then combining Stokes Q and U.
An RM of 6.3 rad\,m$^{-2}$ and a DM of 22\,pc\,cm$^{-3}$ correspond to a line-of-sight averaged magnetic field of approximately 0.4\,$\mu$G, which is in agreement with the average magnetic field in the Galactic interstellar medium in the direction of J1634+44 \citep{2023NatAs...7.1295R}.

In a few of the linearly polarised pulses, we see a variation in the Stokes\,V fraction as a function of frequency. This variation is much slower than the clear Faraday rotation seen in Stokes\,Q and U, and could simply be intrinsic variation in the pulse polarisation with frequency. It could also be a consequence of generalised Faraday rotation in a relativistic plasma, where the rotation no longer just affects Stokes\,Q and U, but also Stokes\,V \citep[e.g.][]{1969JETP...29..578S,1998PASA...15..211K}. If we treat this variation as relativistic Faraday rotation, the corresponding relativistic rotation measure would be $\sim0.1\pm0.05$\,rad\,m$^{-3}$. This is not inconsistent with relativistic Faraday rotation in the magnetosphere of a compact object, but the uncertainties are too large to say anything more.

Curiously, the polarisation properties of the pulses from J1634+44 seem to be correlated with the phase of the pulse, as can be seen in Figure\,\ref{fig:lc_wrapped}. The pulses that arrive after three periods are always strongly circularly polarised, while all but one of the pulses that arrive after two periods are strongly linearly polarised.

\subsection{Occurrence rate}
J1634+44 was found serendipitously in the LoTSS DR2 survey, specifically in the Stokes V transient search that will be described in detail in Bloot et al. (in prep). This search consisted of 841 pointings, each exposed for 8 hours, with a FWHM of 2 degrees. No other LPTs were found above an 8 sigma threshold. Considering the average sensitivity of the observations and the total sky coverage, we conclude that the occurrence rate of bursts above 12\,mJy when integrated over 4 minutes in Stokes\,V is 0.9$^{+2.2}_{-0.8} \times 10^{-4}$ per square degree per epoch, where the error bars are based on Poisson bounds for low-number statistics \citep{1986ApJ...303..336G}. When LoTSS is complete, we therefore expect to find $4^{+9}_{-3}$ of these sources. However, we note that this prediction assumes that the distribution of these sources is homogenous across the sky, and that the noise is constant across the entire survey. If there is an overdensity of LPTs on the Galactic plane, this estimate is a lower limit to the number of detections we can expect, although taking into account the expected higher noise near the Galactic plane may cancel out this effect.

\section{Multi-wavelength photometry}
\label{sec:wl}
\begin{figure*}
    \centering
    \includegraphics[width=0.95\linewidth]{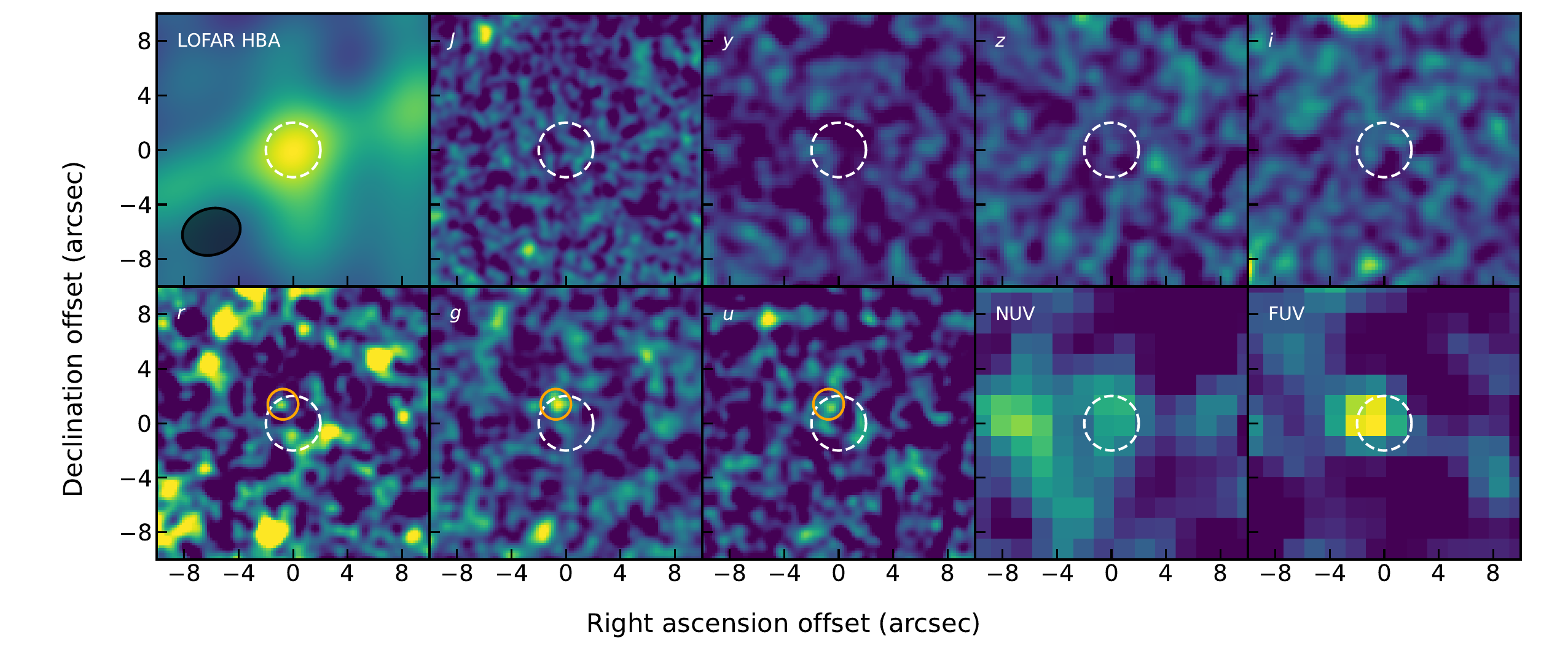}
    \caption{Cut-out images of the location of J1634+44 at the various wavelengths where it has been observed. The first row contains, from left to right, the Stokes\,V LOFAR image of the brightest pulse, the CFHT $J$-band image, and the $y$, $z$ and $i$ images from Pan-STARRS. The second row contains, again from left to right, the UNIONS images in the $r$, $g$ and $u$-bands, the GALEX MIS NUV image, and the GALEX MIS FUV image. In each image, a dashed white line shows the 2" positional uncertainty region around the measured position of J1634+44. The FWHM of the beam in the LOFAR image is indicated by the shaded black ellipse. In the UNIONS images, an orange circle shows the location of the detections.}
    \label{fig:all_images}
\end{figure*}

J1634+44 is located $\sim$40$\degree$ off the Galactic plane, making it an ideal target for multi-wavelength follow-up due to the low extinction along the line of sight. We searched for a counterpart from $y$-band to the far-ultraviolet (FUV) in survey data. We also acquired a deep $J$-band exposure with UKIRT. We include cut-outs of all images in Fig.\,\ref{fig:all_images}.

\subsection{Survey data}
\subsubsection{Pan-STARRS}
J1634+44 was covered in Pan-STARRS \citep{2016arXiv161205560C,2020ApJS..251....3M,2020ApJS..251....4W,2020ApJS..251....5M,2020ApJS..251....6M,2020ApJS..251....7F}, with no detections in the stacked images. To determine formal upper limits on the magnitude of J1634+44 in the PAN-STARRS observations, we use the cut-outs of the stacked images in the $g$, $r$, $i$, $z$, and $y$ filters. We determine the FWHM of the PSF in the images through \texttt{PSFex} \citep{2011ASPC..442..435B}, and use it as the aperture diameter to measure the flux. We apply an aperture correction based on the PSF and determine the $3$\,sigma AB magnitude limit in each filter, listed in Table\,\ref{tab:limits}.

\subsubsection{GALEX}
The GALEX Medium Imaging Survey \citep[MIS,][]{2005ApJ...619L...1M} surveyed the Northern sky in both the FUV and NUV bands, providing source catalogues and images. J1634+44 was covered in both NUV and FUV, with an exposure time of 1594\,s. The NUV observation resulted in a 3\,$\sigma$ magnitude limit of 23.6 at the location of J1634+44. The FUV image, however, revealed a 5.2\,$\sigma$ source at the J2000 position of ($248^\circ.6250697, 44^\circ.8359621$), which agrees, within uncertainties, with the radio position of J1634+44. The source has an FUV AB magnitude of $23.8\pm0.2$. It was not listed in the MIS catalogue originally, as it is just barely below the threshold for detection in the survey source finder configuration.

\subsubsection{UNIONS}
The field around J1634+44 was observed in the Ultra-violet Near Infrared Optical Northern Survey \citep[UNIONS;][]{gwyn2025unionsultravioletnearinfraredoptical}, in the $u$ and $r$ filters as a part of the  Canada-France Imaging Survey \citep[CFIS,][]{2017ApJ...848..128I} with CFHT/MegaCam, and in the $g$-band as part of the Waterloo-Hawaii IfA $g$-band Survey with Subaru. 
We used the deep stack images from UNIONS in the $u$, $g$, and $r$ filters and again extract the FWHM using \texttt{PSFEx} to use as the aperture diameter. We detect a source with a significance of $3.5\sigma$, $3.2\sigma$, and $3.9\sigma$ respectively in the $u$, $g$ and $r$ filters, within 2\arcsecfrac0 from the position of J1634+44 in the radio observations.
Each individual detection is not very significant, but the source is detected in all three filters, at the same location, making the detection more convincing. The source agrees with the position of the GALEX FUV source within a PSF width, further increasing our confidence in the detection.
The AB magnitudes of the detections in the three filters are listed in Table\,\ref{tab:limits}.

\subsubsection{ROSAT}
J1634+44 was observed as a part of the ROSAT All Sky Survey \citep{1999A&A...349..389V} but was not detected. We obtained an upper limit from the Upper Limit server\footnote{\url{http://xmmuls.esac.esa.int/upperlimitserver/}}, resulting in a $3\sigma$ upper limit on the flux of $2.24\times10^{-13}$\,ergs\,cm$^{-2}$\,s$^{-1}$ between 0.2-2.0\,keV.

\subsection{UKIRT}
We obtained deep $J$-band images of the field containing J1634+44 using the near-infrared Wide Field Camera \citep[WFCAM;][]{2007A&A...467..777C} on UKIRT. WFCAM consists of four $2048\times2048$ Rockwell Hawaii-II (HgCdTe) infrared arrays, each with a field of view of $13.\!'65 \times 13.\!'65$. We placed the co-ordinates of J1634+44 at the centre of Camera 3 (WFCAM's northeast array).

All UKIRT observations were performed in service (queue) mode. Observations took place on 14 nights between
\hbox{2024 September 28 UT} through \hbox{2024 October 20 UT}. For each observation, we obtained three sets of nine 60~sec exposures taken in UKIRT's 3$\times$3 large microstepping sequence, for a total of $27\times60=1620$~sec integration time per night.
These were usually the first observation of the night and often occurred at least partly in $12^{\circ}$ twilight. Seeing ranged from 0.7--1.1 arcsec, with thin cirrus or clear conditions. 

All data were reduced in standard fashion by the WFCAM Cambridge Astronomical Survey Unit (CASU) pipeline \citep{2004SPIE.5493..411I,2007MNRAS.379.1599L}, including detrending using facility dark and flat-field images as well as photometric and astrometric calibration. The CASU pipeline aligned the 3$\times$3 microstepped frames from each set of nine exposures and placed the images on an interwoven grid such that each 0.4" pixel was represented by a 0.133" pixel centered at the same position. Following our observations, CASU staff stacked the interwoven images from each night into a single image with a nominal exposure time of 6.3 hours.
We do not detect a source at the location of J1634+44, resulting in a 3$\sigma$ upper limit of 24.7 in AB magnitudes when using an aperture size based on the FWHM from \texttt{PSFex}.

\begin{table}
\def\arraystretch{1.25}
\centering 
\caption{Multi-wavelength follow-up results.}
    \begin{tabular}{lrr}
    \hline \hline
    Filter & $m_{\mathrm{AB}}$ & 3$\sigma$ $m_{\mathrm{AB}}$ \cr
    \hline \hline
    \textit{GALEX MIS} & \cr
    \hline
    FUV & 23.81$_{-0.19}^{+0.23}$ & 24.40 \cr
    NUV & - & 23.58 \cr
    \hline
        \textit{UNIONS} & \cr
    \hline
    $u$ & 25.07$_{-0.27}^{+0.36}$ & 25.25 \cr
    $g$ & 25.29$_{-0.29}^{+0.39}$  & 25.39 \cr
    $r$ & 25.61$_{-0.25}^{+0.32}$  & 25.90 \cr
    \hline
    \textit{Pan-STARRS} & \cr
    \hline
    $g$ & - & 24.40 \cr
    $r$ & - & 24.20  \cr
    $i$ & - & 24.18 \cr
    $z$ & - & 22.96 \cr
    $y$ & - & 21.65 \cr
    \hline
    \textit{UKIRT} & \cr
    \hline
    $J$ & - & 24.70 \cr 
    \hline
    \end{tabular}%
    \tablefoot{AB magnitudes with 1\,$\sigma$ error bars for filters with detections, as well as 3$\sigma$ limits for all images, for J1634+44.}
    \label{tab:limits}
\end{table}

\section{Progenitor system of J1634+44}
\label{sec:system}
Although the progenitor system of most LPTs remains unknown, they are thought to be either magnetars or white dwarfs. In two systems, ILT\,J1101+5521 and GLEAM-X\,J0704–37, a main-sequence star was detected at the location of the LPT, with a blue excess and radial velocity shifts indicating the likely presence of a white dwarf \citep{iris, hurley-walker24, 2025arXiv250103315R}.

We detect J1634+44 in the FUV, as well as in the $u$, $g$ and $r$ filters (see Tab.\,\ref{tab:limits}). Magnetars are not expected to produce bright FUV or blue emission, nor are they expected to be found at such a high Galactic latitude. We therefore conclude that a magnetar origin for J1634+44 is unlikely.

Instead, the detections in GALEX and UNIONS, combined with the limits in the other filters, suggest a white dwarf as part of the progenitor system of J1634+44. As the uncertainties on the detected flux are quite large, we cannot uniquely determine the properties of the white dwarf, even with four filters. Instead, we  determine the range of properties that agree with the measurements. We use the \citet{koester} synthetic white dwarf spectra, in combination with cooling models\footnote{\url{http://www.astro.umontreal.ca/~bergeron/CoolingModels}} based on \citet{1995PASP..107.1047B}, \citet{2006AJ....132.1221H}, \citet{2006ApJ...651L.137K}, \citet{2011ApJ...737...28B}, \citet{2011ApJ...730..128T}, \citet{2018ApJ...863..184B}, and \citet{2020ApJ...901...93B} to determine the magnitudes of a white dwarf of a given mass, temperature, log\,$g$, and distance. We assume a pure-hydrogen composition. Assuming a pure-helium composition does not significantly affect our results. We determine log\,$g$ for each combination of the effective temperature and the mass by interpolating the cooling models.

\begin{figure}
    \centering
    \includegraphics[width=0.9\linewidth]{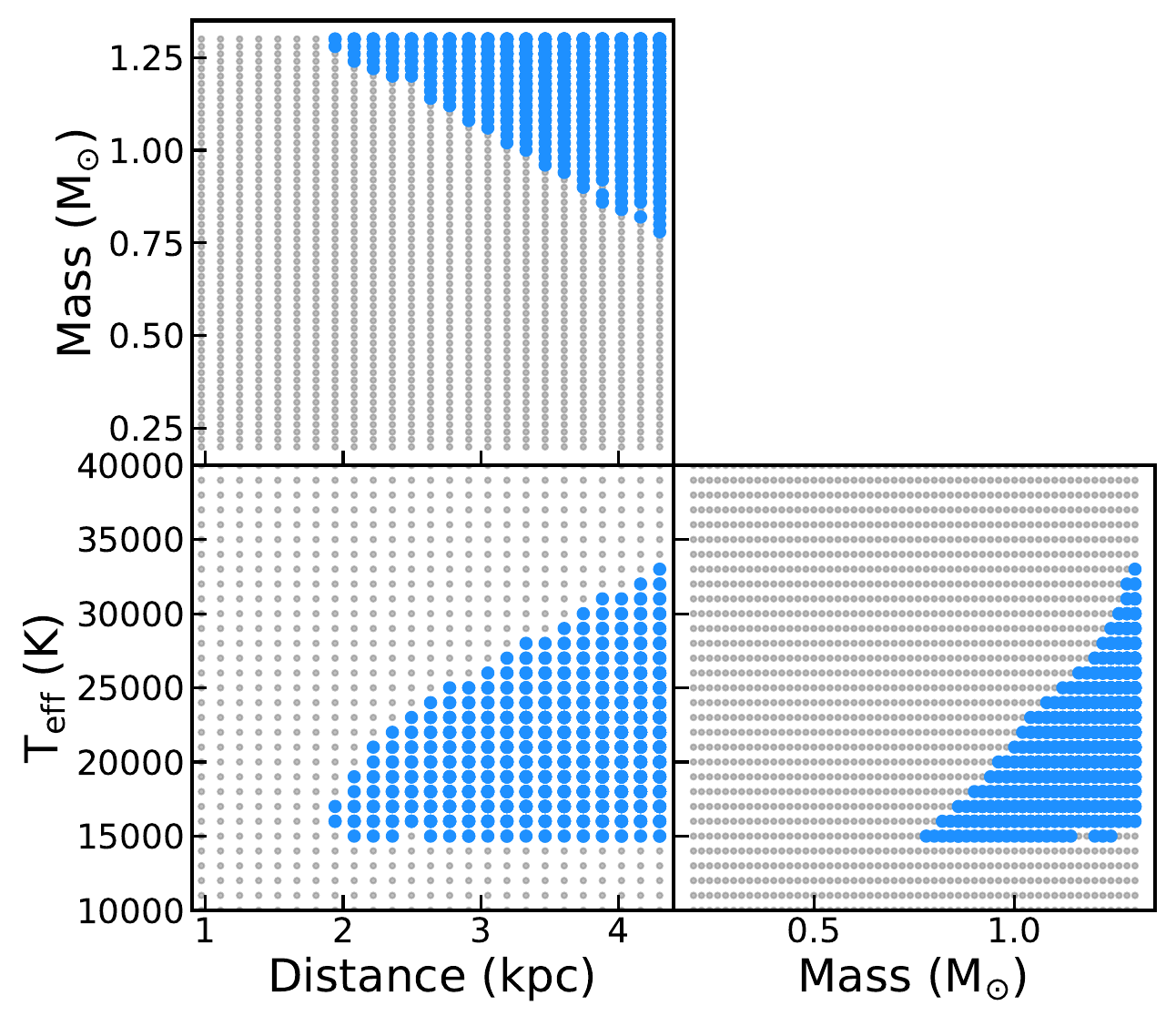}
    \caption{Allowed parameter space for a white dwarf in the J1634+44 system. Each blue point represents a model that agrees with all upper limits and detections within 3$\sigma$. The grey points represent models from the available model grid that do not satisfy our photometric constraints.}
    \label{fig:param_space_wd}
\end{figure}

\begin{figure*}
    \centering
    \includegraphics[width=0.95\linewidth]{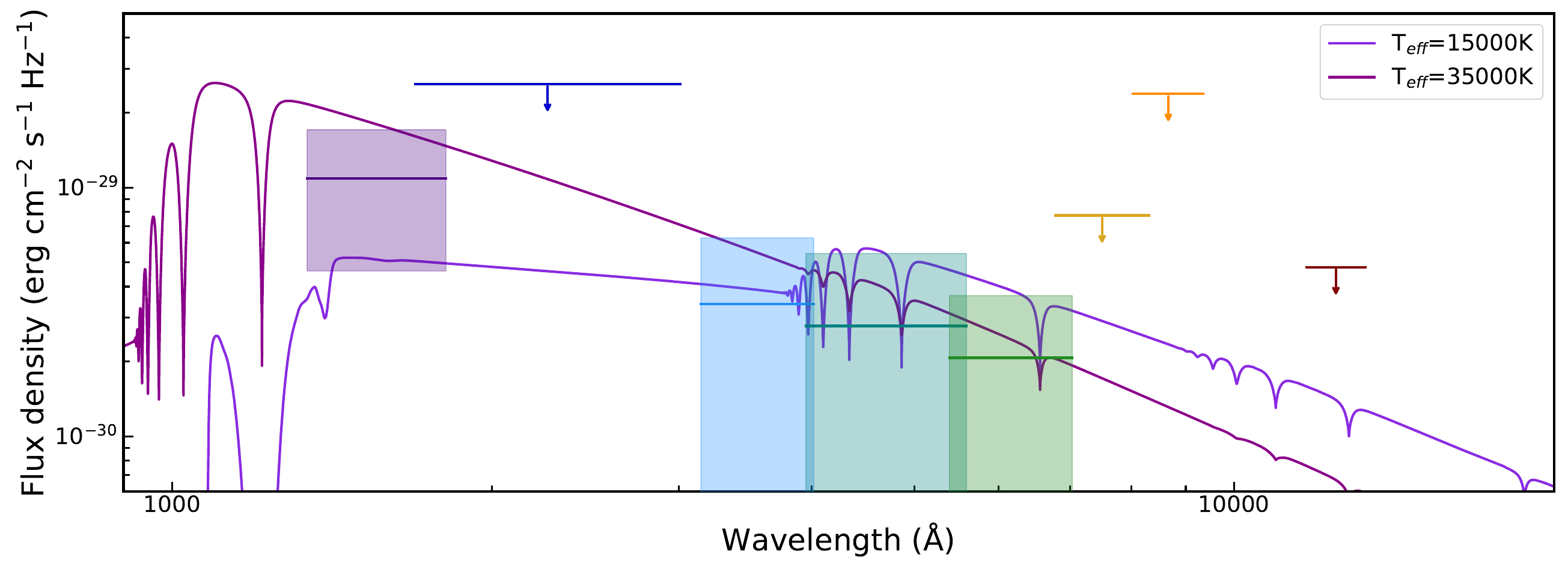}
    \caption{UV to IR spectrum of J1634+44. The horizontal lines represent the limits and detections in the GALEX FUV (purple) and NUV (dark blue) filters, the UNIONS $u$ (light blue), $g$ (teal), and $r$ (green) filters, the Pan-STARRS $i$ (yellow), $z$ (orange) and $y$ (red) filters, and the UKIRT $J$ filter (dark red). The width of the line shows the wavelength coverage of the filter. The lines with arrows are 3$\sigma$ upper limits. The lines with shaded regions are detections, where the shaded regions represent 3$\sigma$ uncertainties. We also plot spectra for an effective temperature of 15000\,K and 35000\,K, which enclose the full range of effective temperatures that satisfy the photometric constraints.}
    \label{fig:spectrum}
\end{figure*}

Combining all data from FUV to IR, we find that the white dwarf has an effective temperature between 15000\,K and 33000\,K, and a distance of more than 1.9\,kpc. Assuming the temperature of the white dwarf is homogeneous across its surface, its mass is 0.78\,\msun or higher. The parameter space is illustrated in Fig.\,\ref{fig:param_space_wd}, and example spectra are plotted in Fig.\,\ref{fig:spectrum}.

If J1634+44 is a binary white dwarf system, we find that the spectral energy distribution is dominated by one of the two components. This component has the same limits as the single white dwarf described above. The properties of the other component are almost completely unconstrained, except for the effective temperature, which must be less than 26000\,K. 

In summary, based on the spectral energy distribution of J1634+44 in the optical and ultraviolet regime, we conclude that the system contains a white dwarf with a mass of 0.78\,\msun or more and an effective temperature between 15000\,K and 33000\,K. The age of this white dwarf is therefore most likely less than 1.5\,Gyr. If there is another white dwarf in the system, its properties are mostly unconstrained, but its effective temperature cannot exceed 26000\,K.

Compared to other radio-emitting white dwarf systems, the white dwarf in J1634+44 has a similar mass, but a relatively high effective temperature. The white dwarf in ILT\,J1101+5521 \citep{iris} has an effective temperature of 4500 to 7500\,K. GLEAM-X\,J0704–37 has a white dwarf effective temperature of $7320_{-590}^{+800}$\,K and a mass of $1.02_{-0.13}^{+0.12}$\,\msun \citep{2025arXiv250103315R}.
AR\,Scorpii \citep{2016Natur.537..374M,2017NatAs...1E..29B} has a mass of 0.8--1.3\,\msun and an effective temperature of 11500\,K. J191213.72-441045.1 \citep{2023NatAs...7..931P} is the most similar to J1634+44, with a mass of 1.2\,\msun and an effective temperature of 13000\,K.

\subsection{(Sub)Stellar companion}
\begin{figure}
    \centering
    \includegraphics[width=0.9\linewidth]{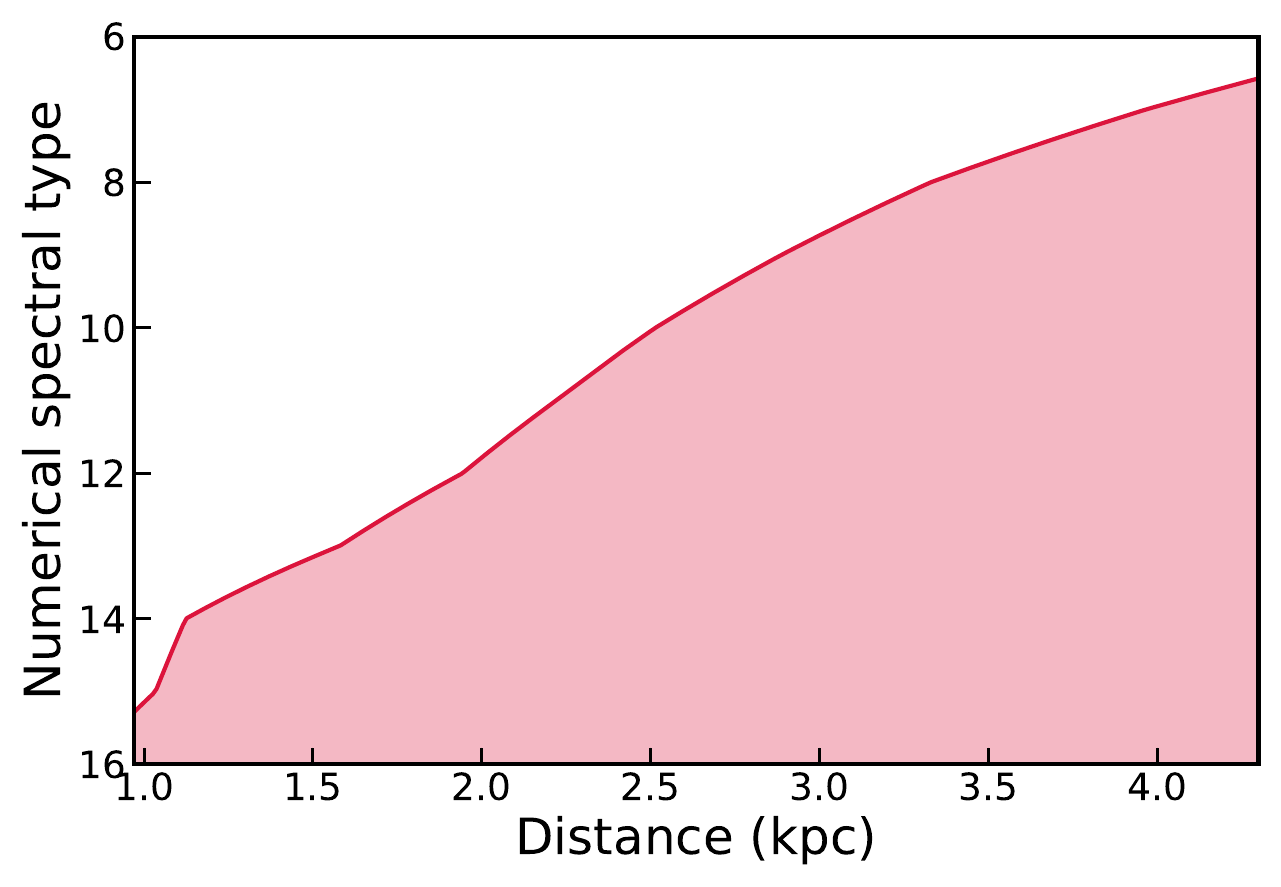}
    \caption{Limit on spectral type of a potential main-sequence star or brown dwarf in J1634+44 as a function of distance. A numerical spectral type of 0 corresponds to M0, 10 to L0, and 20 to T0. The shaded region shows the spectral types that satisfy the photometric constraints in Table\,\ref{tab:limits}.}
    \label{fig:excl_bd}
\end{figure}

White dwarfs are not known to produce bright radio emission in isolation, instead only producing emission when in a binary system with a main-sequence star \citep[e.g.][]{barrett}. Two LPTs have been identified as binaries consisting of a white dwarf and a main-sequence star because a main-sequence star has been detected in the system. We do not detect a main-sequence companion in the J1634+44 system, despite a deep $J$-band integration with a 3$\sigma$ limit of 24.7\,AB mag. 
To determine the nature of a possible main-sequence companion, we use the absolute magnitudes of ultracool dwarfs measured by \citet{2018ApJS..234....1B}. We find that we can rule out a spectral type of M6 at 4.3\,kpc, with the limit as a function of distance plotted in Figure\,\ref{fig:excl_bd}.

The other two LPTs with a white dwarf in the system, ILT\,J1101+5521 and GLEAM-X\,J0704–37, both have a main-sequence star in a tight orbit with the white dwarf. It has been hypothesised that the main-sequence star is necessary to produce the radio emission, either through accretion or through a magnetic interaction \citep{2020ApJ...897....1L,2023PhRvL.130x5201M,2017MNRAS.471L..92K,2016ApJ...823L..28G}. J1634+44 does not completely fit into this paradigm, as the companion has a spectral type of M7 or later.

\subsection{Binary system}
The waiting time analysis clearly shows that the distribution of waiting times between pulses is not random, but instead follows a particular pattern. However, this pattern seems to shift between the 2015--2016 and 2019 observations. Such a shift can be produced as a natural consequence of a binary system with a nearly perfect 5:2 or 5:3 spin-orbit resonance. An essential property of the model is that the instantaneous beaming geometry of the emission is such that it is only visible to us at a certain rotational and orbital phase range. As a particular illustration of the model, we consider a white dwarf and a companion with an orbital period that is $5/2$ times the white dwarf spin period. We assume the white dwarf has a dipolar magnetic field with a non-zero magnetic obliquity, and that it can produce radio emission in a narrow beam along the magnetic axis. We further assume that this radio emission is only produced if the companion is in a certain phase of the orbit, taking a phase range of $-$0.15 to 0.15 as an example. Such an active phase window has been seen before on J191213.72-441045.1 \citep{2023NatAs...7..931P}, where radio pulsations are only produced in certain parts of the orbit.

\begin{figure}
    \centering
    \includegraphics[width=1.0\linewidth,trim={5cm 3cm 5cm 3cm},clip]{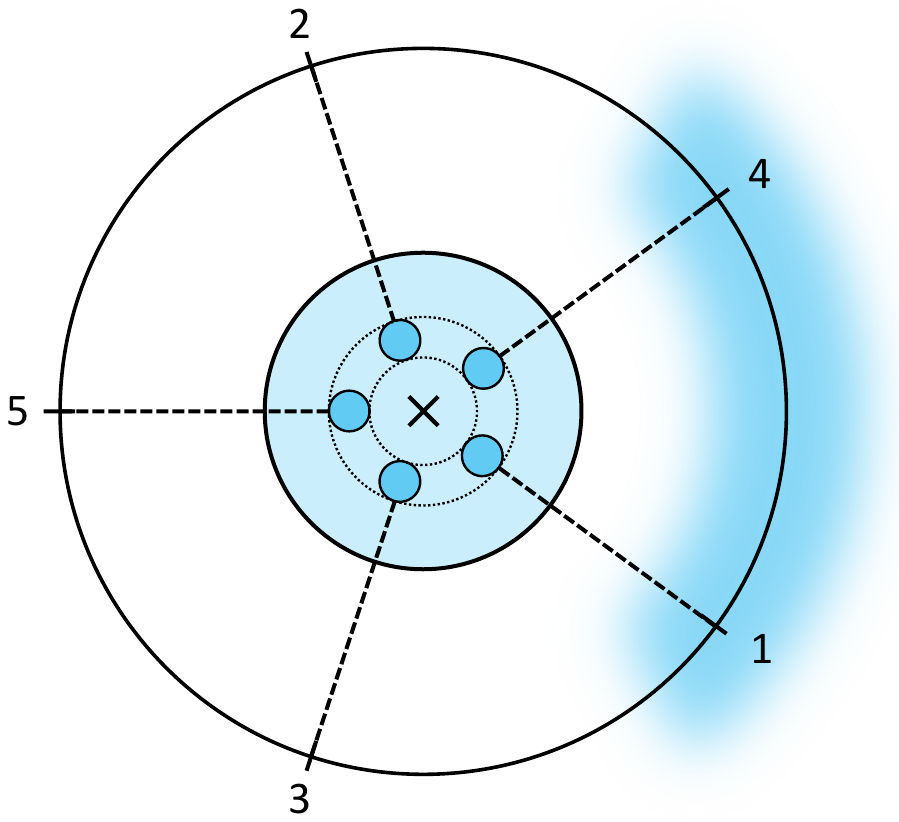}
    \caption{Not-to-scale schematic of the geometry of the spin-orbit resonance model with an active phase window. The white dwarf is shown in light blue. The orbit of the companion is shown by the outer black circle, with the orbital phases corresponding to integer numbers of white dwarf rotations labelled. The cross shows the location of the magnetic axis at a white dwarf rotation phase of zero. The black dashed lines indicate the magnetic field lines connecting the white dwarf and the companion at the corresponding orbital phases, with the blue circles on the white dwarf showing the location of the footpoints of these field lines. The active phase window is shown as a shaded region along the companion orbit.}
    \label{fig:phase_window}
\end{figure}

Combining these assumptions, we will only detect a pulse if the magnetic axis of the white dwarf is aligned with our line of sight (at a rotational phase of 0.0), and when the companion is in the correct phase window (between $-$0.15 and 0.15). Suppose a pulse arrives at a rotational phase of 0.0 and a companion orbital phase of $-$0.1. The next visible pulse arrives three white dwarf rotations later, when the orbital phase of the companion is 0.1. The next pulse after that arrives another two rotations later, when the orbital phase of the companion is $-$0.1 again. We show the geometry of this model in Fig.\,\ref{fig:phase_window}. In this scenario, we expect to see pulses arrive separated by two or three times the white dwarf rotation period in a regular pattern. This explains the emission pattern we see in 2015--2016 and 2019 separately, where the pulses always arrive separated by two, three, or five or more rotation periods, but never by one or four rotation periods.
A 5:3 spin-orbit resonance results in an identical pulse pattern.

The offset between the 2015--2016 and the 2019 data could be explained in two ways: a large period derivative, or a slight misalignment in the spin-orbit resonance. To cause the offset with a period derivative, the period derivative must be large enough to cause the rotational phase to slip by +2 or $-$3 rotations over the period between the 2015--2016 and 2019 observations. With the 841\,s period, the period derivative would therefore have to be $1.0\times10^{-10}$\,s\,s$^{-1}$ or $-1.6\times10^{-10}$\,s\,s$^{-1}$. We do not see such a period derivative in our data.
Instead, we suggest that the system is just slightly off from a perfect spin-orbit resonance. The orbital phase when the white dwarf is aligned with our line of sight would then drift ever so slightly, slowly but surely shifting which rotations produce detectable pulses. If the system produces pulses at rotation one and four out of five at first, a period just slightly shorter than 5/2 times the rotation period of the white dwarf (or slightly longer than 5/3 times the rotation period) will cause the pulses to shift to rotation two and five over time, then three and one, then four and two, and so forth. 

\begin{figure}
    \centering
    \includegraphics[width=0.95\linewidth]{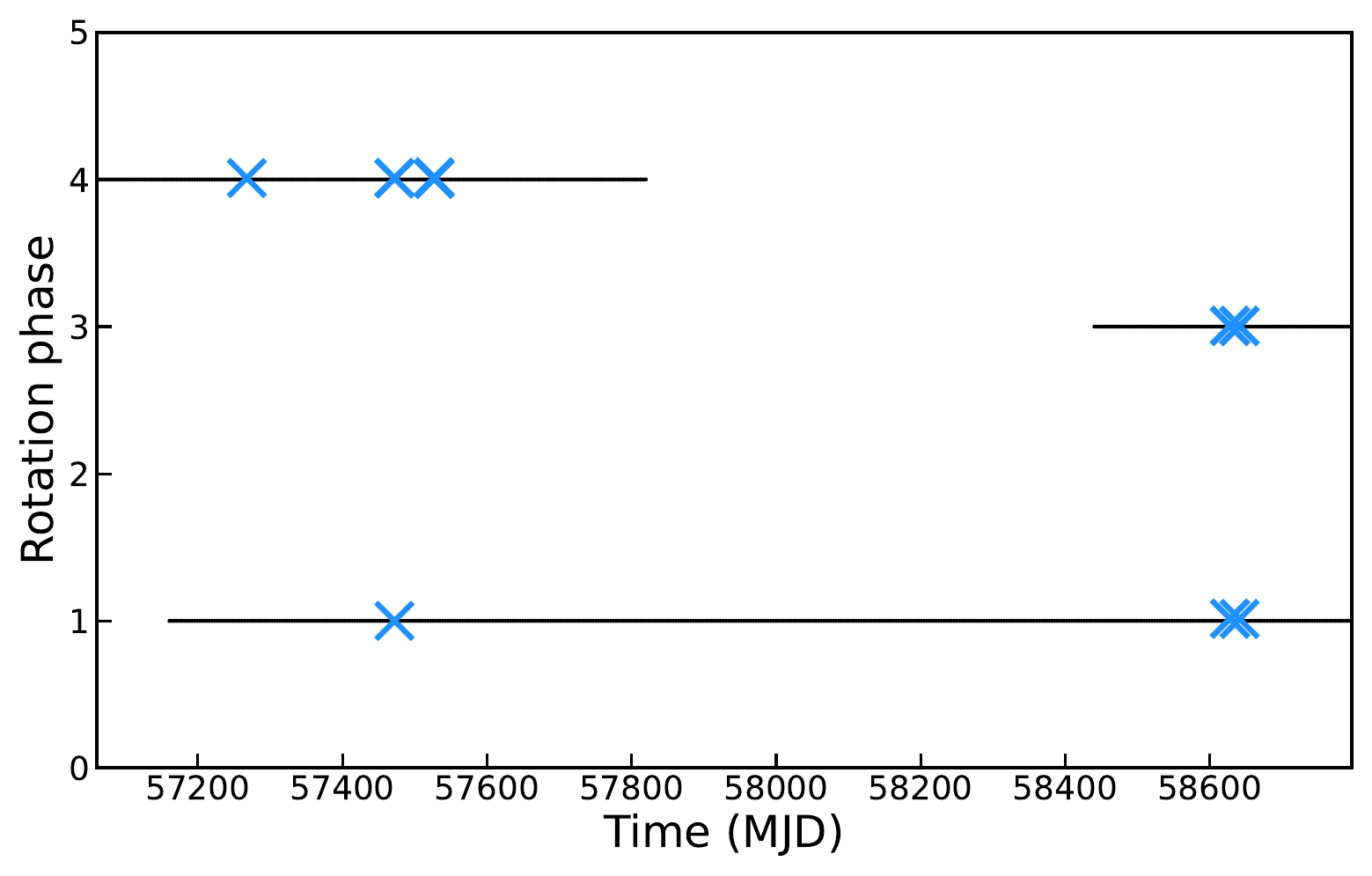}
    \caption{Evolution of the rotation number when the pulses arrive over time. The predicted rotation number is shown in black. We plot the rotation phases of the detected pulses in blue.}
    \label{fig:phases}
\end{figure}

We assume the white dwarf rotation period is the period we measure in the short-period timing solution, $841.24808\pm0.00015$\,s. In our observations, we see pulses arrive at rotation one and four in 2015--2016, and one and three in 2019. Assuming only one change in pulse phases has happened in that time period, and assuming the active phase window covers 0.3 of the orbital phase, we find that the orbit of the companion must be between $2103.1061\pm0.0004$\,s and $2103.1141\pm0.0004$\,s for a 5:2 resonance, and between $1402.0814\pm0.0003$ and $1402.0853\pm0.0003$ for a 5:3 resonance. We show the evolution of the rotation numbers that produce pulses in Fig.\,\ref{fig:phases} for the 5:2 resonance, along with the observed pulse phases.

A requirement for the spin-orbit resonance model is that there is a certain window in the orbital phase where detectable pulses can be produced. We show a diagram of the model in Fig.\,\ref{fig:phase_window}. Out of the five potential pulses indicated in the figure, we require that only pulse one and pulse four are detected. Such a scenario can be arranged in several ways, the first of which is simply through beaming geometry. The pulses last for less than 10\,s, implying that the emission is beamed along a narrow cone that is at most 4.3\degree wide where the line of sight cuts through it. Assuming the pulses are induced by the companion through a magnetic interaction, the emission is likely beamed along the magnetic field line connecting the two bodies. As the footpoint of this magnetic field line changes with the orbital phase of the companion, it is possible to orient the system in such a way that only the pulses from orbital phases in a certain orbital phase window align with the line of sight. The magnetic colatitude of the footpoints must be larger than half the beam width to ensure that not all five pulses are visible. The maximum magnetic colatitude of the footpoint is constrained by the phase offset between the two pulses. The two pulses are produced along field lines separated by 0.2 in phase. In our data, we do not detect a significant offset, implying that the offset is less than 16\,s, or 0.02 in phase. This offset corresponds to a magnetic colatitude of the footpoints of $\lesssim8$\degree. Assuming the magnetic field of the white dwarf is a perfect dipole, the semi-major axis of the orbit of the companion must be larger than 50 white dwarf radii.

Alternatively, an active phase window can be created by a binary interaction, with the companion only interacting with the right pole in a certain phase of the orbit. An example is accretion along the magnetic field lines of the white dwarf, where the material flows towards the pole that is closest to the companion.

We note that we assume that the period measured in the radio data is the rotation period of the white dwarf. It could also be the orbital period of the system, assuming the companion induces emission on the white dwarf only at one specific orbital phase, and that the white dwarf needs to be within a certain rotational phase window to produce detectable pulses. In this scenario, the white dwarf rotation period would be between 2103.10825 and 2103.11229\,s, assuming an orbital period of 841.24808$\pm$0.00015\,s. From the radio data alone, we cannot distinguish between these two scenarios. However, a white dwarf with a rotation period that is much longer than the orbital period seems unlikely, as most white dwarf binaries that are not fully tidally locked consist of a white dwarf with a rotation period shorter than the orbital period \citep[e.g.][]{2025arXiv250308320L} due to mass exchange and conservation of momentum in the final stages of stellar evolution. We therefore proceed with the assumption that the short period of 841\,s is the rotation period of the white dwarf that produces the radio emission, and the longer period is the orbital period of the companion.

From the photometric data, we know that J1634+44 contains at least one white dwarf, and that it does not contain a main-sequence star or a brown dwarf with a spectral type earlier than M7. This leaves three possibilities for the companion: an ultracool dwarf, a white dwarf, or a neutron star.

\subsubsection{Ultracool dwarf companion}
An ultracool dwarf in such a close orbit would likely fill its Roche lobe, so the white dwarf in this system would likely be accreting mass. From the ROSAT non-detection, we constrained the mass accretion rate onto the white dwarf to be $<5\times$10$^{-8}$\,\msun\,yr$^{-1}$ at 4.3\,kpc \citep{1985ApJ...292..535P}. Accretion can spin the white dwarf up, leading to a negative period derivative. The accretion rate required to create the measured period derivative of -$9\times$10$^{-12}$\,s\,s$^{-1}$, assuming accretion via a disc, is $1.6\times10^{-11}$\,\msun\,yr$^{-1}$. Such an accretion rate is well below the accretion rate limit derived from the ROSAT data. 

We consider the plausibility of an ultracool dwarf in such a close orbit with a white dwarf. A system with such a short period has most likely undergone a period of common-envelope evolution, where the companion was engulfed by the white dwarf progenitor as it entered the red giant stage of its life \citep[e.g.][]{1976IAUS...73...75P}. The orbit of the companion would have rapidly shrunk due to the drag of the stellar material. The common-envelope phase is expected to end when the companion reaches the boundary of the convective zone of the red giant, where the energy deposited by the companion can no longer be efficiently transported and the envelope is therefore ejected \citep[e.g.][]{2020MNRAS.497.1895W}. While the companion has likely lost a lot of mass at this stage, even low-mass companions down to a Jupiter mass can eject the envelope \citep{2018ApJ...866...88N}. The separation between the core of the red giant, which will become the white dwarf, and the companion at the time of the envelope ejection is the radius of the boundary of the convective zone. For a core mass of 0.78\,\msun or higher, the radius of the convective boundary ranges from $1.5\times10^{8}$\,m to $2.5\times10^{10}$\,m \citep{2020MNRAS.497.1895W}. Assuming the companion is an ultracool dwarf, the orbital separation of J1634+44 is $\sim5\times10^{10}$\,cm, which is comfortably within the range of possible post-common envelope orbital separations given the white dwarf mass range.

Next, we considered the stability of the orbit against different energy loss mechanisms. The change in orbital period caused by gravitational wave emission is 
\begin{equation}
\label{eq:gw}
    \dot{P}_{\mathrm{GW}} = - \frac{48}{5} \pi^{8/3}\left( \frac{G (M_1 M_2)^{3/5}}{c^3 (M_1+M_2)^{1/5}}\right)^{5/3} \left(\frac{P_\mathrm{orb}}{2}\right)^{-11/3} P_\mathrm{orb}^2,
\end{equation}
where $\dot{P}_{\mathrm{GW}}$ is the change in the orbital period over time, $G$ is the gravitational constant, $M_1$ and $M_2$ are the masses of the two bodies in the system, $c$ is the speed of light, and $P_{\mathrm{orb}}$ is the orbital period of the binary system \citep[][Eq. 18]{1989ApJ...345..434T}.
For a white dwarf with a mass of 1.3\,\msun and an ultracool dwarf with a mass of 0.1\,\msun, $\dot{P}_{\mathrm{GW}}$ is $-1.2\times10^{-12}$\,s\,s$^{-1}$. If the orbit is only affected by gravitational wave emission,this period derivative corresponds to a merger timescale of 80\,Myr.

The orbit can also be affected by ohmic dissipation of currents caused via magnetic induction \citep[e.g.][]{1998ApJ...503L.151L,2002MNRAS.331..221W}. Here the asynchronous orbit of the companion causes unipolar induction to heat up the magnetic white dwarf. The expected energy dissipation in the magnetised white dwarf from induction is 
\begin{equation}
\dot{E}_{\mathrm{UI}}=\frac{1}{\mathcal{R}_{\mathrm{1}}}\left(2R_{\mathrm{2}}\frac{v_{\mathrm{rel}}}{c} B_{\mathrm{1}} \left(\frac{R_\mathrm{1}}{a}\right)^{3}\right)^2 ,
\end{equation}
where $\mathcal{R}_{\mathrm{1}}$ is the effective  resistance of the white dwarf's atmosphere, $R_{\mathrm{2}}$ is the radius of the ultracool dwarf, $R_{\mathrm{1}}$ is the radius of the white dwarf, $v_{\mathrm{rel}}$ is the orbital velocity of the ultracool dwarf in a frame corotating with the white dwarf, $B_{\mathrm{1}}$ is the surface dipole magnetic field strength of the white dwarf, and $a$ is the semi-major axis of the orbit \citep[][Sec.\,2]{2002MNRAS.331..221W}. The corresponding period derivative of the orbital period, assuming all of the energy is extracted from the orbital energy of the system, is given by
\begin{equation}
    \dot{P}_{\mathrm{UI}}=\frac{\dot{E}_{\mathrm{UI}} P_{\mathrm{orb}}}{g(P_{\mathrm{orb}}) (1-\alpha)},
\end{equation}
where $\alpha$ is the ratio of the white dwarf spin period and the orbital period. $g(P_{\mathrm{orb}})$ is given by
\begin{equation}
    g(P_{\mathrm{orb}})=-\frac{1}{3}\left(\frac{q^3}{1+q} G^2 M_{1}^5 \left(\frac{2\pi}{P_\mathrm{orb}}\right)^{2}\right)^{1/3} \left( 1-\frac{6}{5}(1+q)f(P_{\mathrm{orb}})\right),
\end{equation}
where $q$ is the ratio of $M_1$ and $M_2$, and 
\begin{equation}
    f(P_{\mathrm{orb}})=\left(\frac{4\pi^2 R_2^3 }{G M_1 (1+q)P_{\mathrm{orb}}^2} \right)^{2/3}.
\end{equation}
For a white dwarf with a mass of 1\,\msun, a magnetic field strength of 1\,kG, an electron temperature of 10$^4$\,K, and an ultracool dwarf with a mass of 0.1\,\msun and a radius of 0.12\,R$_{\odot}$, $\dot{P}_{\mathrm{UI}}$ is $-1.5\times10^{-12}$\,s\,s$^{-1}$. If the magnetic field of the white dwarf is 1\,MG instead, $\dot{P}_{\mathrm{UI}}$ is $-2.5\times10^{-6}$\,s\,s$^{-1}$, which would lead to the two objects merging in only 77\,yr. We therefore conclude that a binary of a white dwarf and an ultracool dwarf can be stable in this orbit for a few tens of Myr, provided the magnetic field strength of the white dwarf is $\lesssim10^4$\,G.

In cataclysmic variable (CV) systems, which consist of a white dwarf and a companion main-sequence star or brown dwarf that is accreting onto the white dwarf, the orbit is expected to shrink until the companion overflows its Roche lobe, at which point the orbital period starts increasing. CVs are therefore expected to have a minimum orbital period, which has been estimated to be at 40\,min \citep{2018ApJ...866...88N}. J1634+44, with its 35\,min period, is below this period minimum, meaning that its orbital period should be increasing due to accretion. The rate of change of the orbital period as a consequence of accretion from an ultracool dwarf can be estimated according to \citet{2018ApJ...866...88N} as \begin{equation}
    \dot{P}_{\mathrm{acc}}\approx\frac{3}{11} \left(\frac{352 (2\pi)^{8/3} G^{5/3} (M_{\mathrm{UCD}}+M_{\mathrm{WD}})^{5/3} M_{\mathrm{UCD}} }{10 c^5 P_{\mathrm{orb}}^{5/3} M_{\mathrm{WD}}}\right).
\end{equation}
For a white dwarf mass of 1.2\,\msun and an ultracool dwarf mass of 0.1\msun, $\dot{P}_{\mathrm{acc}}$ is $7.2\times10^{-13}$\,s\,s$^{-1}$, which is comparable in magnitude to $\dot{P}_{\mathrm{GW}}$ and $\dot{P}_{\mathrm{UI}}$. We therefore conclude that it is possible to create a binary system with an ultracool dwarf and a white dwarf in a $\sim$35\,min period and keep it in this orbit, as all effects acting to increase or decrease the orbital period are small and can compensate for each other to a certain extent.

\subsubsection{White dwarf companion}
The photometry we have available on the system is also consistent with the companion to J1634+44 being another white dwarf, making J1634+44 a double white dwarf (DWD) binary. With two white dwarfs in the system, accretion is unlikely. This makes the negative period derivative measured in this work and by \citet{adam} difficult to explain. While a while dwarf can have a negative period derivative early in its life, when it is still contracting, the period derivative from dynamical contraction at the ages that agree with the photometric data are too small by at least an order of magnitude \citep{2018MNRAS.474.2750P}. However, it is possible that the dynamical contraction is affected by binary interactions, possibly extending the phase where the white dwarf is spinning up. 

A DWD binary with such a short orbit is likely to have gone through a common-envelope phase. While we do not know which component is older, it seems likely that the less massive white dwarf is the younger of the two. As the mass of the less massive white dwarf is unconstrained, the range of expected orbital separations after the common envelope phase is larger than for an ultracool dwarf companion, ranging from $7\times10^{7}$\,m to $3.5\times10^{10}$\,m \citep{2020MNRAS.497.1895W}. For all combinations of masses that agree with the photometric data, the orbital separation implied by the orbital period in our binary fits fall in this range.

To determine how this system will evolve, we can again evaluate $\dot{P}_{\mathrm{GW}}$ and $\dot{P}_{\mathrm{UI}}$. From Eq.\,\ref{eq:gw}, $\dot{P}_{\mathrm{GW}}$ must be between $-1.8\times10^{-12}$\,s\,s$^{-1}$ and $-1.2\times10^{-11}$\,s\,s$^{-1}$ for the range of masses that agree with the photometric data. If gravitational wave emission is the only process affecting the orbit of the system, it would merge in 8--60\,Myr.

Assuming one of the white dwarfs is unmagnetised, we can determine the energy loss through unipolar induction. We assume an electron temperature of 10$^4$\,K and a magnetic field strength for the magnetic white dwarf of 1\,kG. Assuming the magnetic white dwarf is the massive component, $\dot{P}_{\mathrm{UI}}$ is higher than $-5.5\times10^{-14}$\,s\,s$^{-1}$. If we instead assume the less massive component is the magnetic white dwarf, $\dot{P}_{\mathrm{UI}}$ is higher than $-1.6\times10^{-12}$\,s\,s$^{-1}$. Increasing the magnetic field strength to 1\,MG, we find that $\dot{P}_{\mathrm{UI}}$ is between $-3\times10^{-15}$\,s\,s$^{-1}$ and $-5.5\times10^{-8}$\,s\,s$^{-1}$ if the magnetic white dwarf is the massive component, or between $-3\times10^{-15}$\,s\,s$^{-1}$ and $-1.6\times10^{-6}$\,s\,s$^{-1}$ if the magnetic white dwarf is the less massive component.

Both gravitational wave emission and unipolar induction can decrease the orbit of J1634+44, potentially disrupting the spin-orbit resonance. However, the white dwarf producing the radio emission has a measured negative period derivative, which can keep the system in resonance longer. We conclude that the orbit and even the resonance in a DWD binary can be stable over long periods of time, as long as the magnetic field of the white dwarfs is not too strong ($\lesssim10^4$\,G).

We note that the companion to the white dwarf in J1634+44 could also be a neutron star. A neutron star would cause a larger energy loss through gravitational wave emission, which makes the system less stable. Otherwise, the considerations are the same as for a DWD binary system.

\section{Radio emission process}
\label{sec:em}
The radio emission from J1634+44 shows several curious properties. The pulses are bright and strongly variable in flux density. They show a clear periodicity at $841$\,s, but with pulses only arriving after two or three rotations. The pulses that arrive after two rotations ("pulse 1" in the rest of this section) are almost 100\% circularly polarised, while all but one of the pulses that arrive after three rotations ("pulse 2" in the rest of this section) are strongly linearly polarised. Most LPTs have consistent polarisation properties, where the pulses are strongly linearly polarised with some circular polarisation. The exception to this rule is ASKAP\,J193505.1+214841.0 \citep{2024NatAs...8.1159C}, which has emission states with different polarisation properties. However, ASKAP\,J193505.1+214841.0 does not seem to change polarisation states within a few periods, whereas J1634+44 has different emission states within five rotations.
The short duration of the pulses from J1634+44, combined with the high polarisation fraction of the bursts, confidently rules out an incoherent emission mechanism. 

If J1634+44 has an ultracool dwarf companion, the system could be similar to radio-emitting CVs. CVs are known to produce strongly circularly polarised pulses \citep[e.g.][]{barrett}, which are ascribed to emission from the electron-cyclotron maser instability (ECMI) from the main-sequence star in the system. While J1634+44 produces strongly circularly polarised pulses, it also produces strongly linearly polarised pulses. ECMI in general produces elliptically polarised emission \citep[e.g.][]{dulk1985} that circularises due to propagation effects. As such, 100\% linearly polarised ECMI emission has not been observed. We can therefore rule out classical (non-relativistic) ECMI as the emission process powering J1634+44.

A potential origin for the radio emission from LPTs is relativistic ECMI, as proposed by \citet{qu_zhang}. In this model, the energy is provided through the unipolar inductor effect due to a companion with a magnetic field moving through the magnetosphere of a white dwarf. The total electric power dissipation rate of a binary system can be estimated according to Equation\,11 in \citet{qu_zhang}. Assuming an ultracool dwarf companion, with a mass of 0.1\,\msun and a radius of 0.1\,R$_{\odot}$, and a white dwarf magnetic field strength of 1\,kG, we find that the unipolar inductor effect can provide the energy required to produce the brightest pulses ($8.8\times10^{21}$\,erg\,s$^{-1}$ at 4.3\,kpc) given the ratio of the white dwarf rotation period to the orbital period of $\sim$0.4. Relativistic ECMI can produce emission that is strongly linearly polarised or strongly circularly polarised, depending on the viewing angle compared to the magnetic field. 

The luminosity and the periodic nature of the emission could also be explained by an emission process similar to the emission mechanism that powers pulsars. In this framework, however, the polarisation properties of the pulses are puzzling. Pulsars are known to sometimes produce strongly linearly polarised pulses, and in rare cases circular polarisation fractions up to 60\% \citep[e.g.][]{1998MNRAS.301..235G,han2020, vlotss}, but are not known to produce pulses with 100\% circular polarisation. Curvature radiation, in theory, can produce strong circular polarisation, but producing 100\% circular polarisation requires specific geometries. The polarisation fraction measured from a curvature radiation source strongly depends on where the sight line cuts the beam, as shown by \citet{wang2012}. In theory, it must be possible to produce a beam pattern where one can cut the beam in such a way as to only see strong circular polarisation, with minimal linear polarisation. Similarly, changing the angle at which the sight line passes through this beam can result in a purely linearly polarised pulse profile. We therefore suggest the possibility that the difference in polarisation between pulse 1 and pulse 2 is caused by the two emission beams crossing the sight line with different impact angles. However, in this scenario, one would naively assume that the total intensity of emission for the pulses on pulse 2 must be higher, as the sight line cuts the beam closer to the maximum power point. While we do not have enough detections to test whether this is the case, we detect a roughly equal number of pulses 1 and 2 in 2019, implying that if such an effect exists, it is not very significant.

Alternatively, the different polarisation states could be caused by propagation effects. We assume the white dwarf must be at the same rotation phase for every pulse. If the different polarisation states are caused by propagation effects, these effects must therefore be dependent on the orbital phase of the companion.
As an example, we assume the emission is produced with 100\% linear polarisation. As the emission propagates through a magnetised plasma, it is possible for one of the two magnetoionic modes to be fully absorbed, for example through cyclotron absorption \citep{1982ApJ...259..844M}. 
In a non-relativistic medium, the two modes are circular. Fully absorbing one mode would therefore result in emission that is 100\% circularly polarised. As the polarisation seems to depend on the orbital phase of the companion, the absorption process must be modulated by the orbit. Such a scenario can be created by having the absorption happen in the magnetosphere of the companion, or in the accreting material falling onto the white dwarf. However, both options require a specific geometry where the companion's magnetosphere or accreting material eclipses the white dwarf.

If J1634+44 has an ultracool dwarf companion, the difference in polarisation properties can be caused by the variable accretion onto the visible pole. If the pulse is produced when accretion has started only shortly before, the density in the emitting region is still low, allowing a linearly polarised pulse to escape the magnetosphere relatively unscathed. If the pulse is emitted near the end of the accretion phase, the density in the magnetosphere is much higher, potentially blocking one of the two modes. Alternatively, the change in density can affect whether the modes are linear or circular, resulting in pulses being produced with different polarisation properties.

In summary, the radio emission from J1634+44 is likely produced either through a mechanism analogous to that operating on pulsars or through relativistic ECMI. The different polarisation states can be caused by a difference in intrinsic polarisation, possibly caused by the beaming geometry, or by a difference in propagation effects.

\section{Conclusions}
\label{sec:conc}
We report the detection of a new long-period transient, ILT\,J163430+445010, which produces strongly circularly and strongly linearly polarised pulses. The dispersion measure of 22.5$\pm$5.5 pc cm$^{-3}$ indicates that its distance is 1.0 to 4.3\,kpc. The rotation measure of 6.3 rad\,m$^{-2}$ agrees with Faraday rotation by the interstellar medium at this distance range. Follow-up observations and archival data show no detections at infrared wavelengths, but the system is tentatively detected in both GALEX and UNIONS, with an FUV AB magnitude of 23.8 and $u$, $g$, and $r$ AB magnitudes of 25.1, 25.3, and 25.6 respectively. We conclude that J1634+44 is most likely a white dwarf with an effective temperature between 15000\,K and 33000\,K, and a mass over 0.78\,\msun if we assume the white dwarf has a homogeneous surface temperature.  However, deeper optical and ultraviolet observations are required to confirm the white dwarf and constrain its properties. From the deep $J$-band exposure, we can rule out a main-sequence or brown dwarf companion with a spectral type earlier than M7.

The pulses detected on J1634+44 are periodic with a period of 841\,s. However, the pulses do not arrive randomly, instead following a specific pattern of producing two pulses every five periods, which shifts over time. We show that such a pattern can be explained with a binary system in a nearly perfect 5:2 or 5:3 spin-orbit resonance, with a companion orbiting the white dwarf with a period of 2103\,s or 1402\,s. The companion could be an ultracool dwarf, a second, colder white dwarf, or a neutron star. An ultracool dwarf companion could explain the negative spin period derivative observed in J1634+44, while this is more difficult in a double white dwarf system. To uniquely determine the nature of the companion, deep photometric observations at both UV and infrared wavelengths are required, as well as long-term radio monitoring to confirm the pulse pattern.

The emission mechanism powering the bright and strongly polarised pulses must be coherent and able to produce strong polarisation. We suggest that the emission mechanism is either relativistic ECMI or the emission mechanism that powers pulsars, both of which can produce strongly polarised pulses. The variable polarisation properties can be caused by an intrinsic change in polarisation with viewing angle or a by propagation effect that varies along the orbit of the companion.

J1634+44's polarisation properties and pulse timing make it unique even among the small number of known LPTs. However, when LoTSS is complete, we expect to find on the order of a few more objects similar to J1634+44.

\subsection*{Data availability}
The LOFAR data supporting this work is publicly available as part of LoTSS DR2 at \url{https://repository.surfsara.nl/collection/lotss-dr2} and in the Long-Term Archive at \url{https://lta.lofar.eu/}. The GALEX images are available in the GALEX data archive hosted at \url{https://archive.stsci.edu/missions-and-data/galex}. The PAN-STARRS images were retrieved from the image cut-out server at \url{http://ps1images.stsci.edu/cgi-bin/ps1cutouts}.

\begin{acknowledgements}
\noindent SB and HKV acknowledge funding from the Dutch research council (NWO) under the talent programme (Vidi grant VI.Vidi.203.093). HKV acknowledges funding from the ERC starting grant `Stormchaser' (grant number 101042416). JRC acknowledges funding from the European Union via the European Research Council (ERC) grant Epaphus (project number 101166008). TJD acknowledges support from UKRI STFC AGP grant ST/W001209/1.

This research made use of NASA's Astrophysics Data System, the \textsc{IPython} package \citep{PER-GRA:2007}; \textsc{SciPy} \citep{scipy}; \textsc{Matplotlib}, a \textsc{Python} library for publication quality graphics \citep{Hunter:2007}; \textsc{Astropy}, a community-developed core \textsc{Python} package for astronomy \citep{2013A&A...558A..33A}; and \textsc{NumPy} \citep{van2011numpy}. 

This paper is based on data obtained with the LOFAR telescope (LOFAR-ERIC) under project codes LC4\_034 and LT10\_010. LOFAR (van Haarlem et al. 2013) is the Low Frequency Array designed and constructed by ASTRON. It has observing, data processing, and data storage facilities in several countries, that are owned by various parties (each with their own funding sources), and that are collectively operated by the LOFAR European Research Infrastructure Consortium (LOFAR-ERIC) under a joint scientific policy. The LOFAR-ERIC resources have benefited from the following recent major funding sources: CNRS-INSU, Observatoire de Paris and Université d'Orléans, France; Istituto Nazionale di Astrofisica (INAF), Italy; BMBF, MIWF-NRW, MPG, Germany; Science Foundation Ireland (SFI), Department of Business, Enterprise and Innovation (DBEI), Ireland; NWO, The Netherlands; The Science and Technology Facilities Council, UK; Ministry of Science and Higher Education, Poland.
This publication is part of the project LOFAR Data Valorization (LDV) [project numbers 2020.031, 2022.033, and 2024.047] of the research programme Computing Time on National Computer Facilities using SPIDER that is (co-)funded by the Dutch Research Council (NWO), hosted by SURF through the call for proposals of Computing Time on National Computer Facilities. 

The Pan-STARRS1 Surveys (PS1) and the PS1 public science archive have been made possible through contributions by the Institute for Astronomy, the University of Hawaii, the Pan-STARRS Project Office, the Max-Planck Society and its participating institutes, the Max Planck Institute for Astronomy, Heidelberg and the Max Planck Institute for Extraterrestrial Physics, Garching, The Johns Hopkins University, Durham University, the University of Edinburgh, the Queen's University Belfast, the Harvard-Smithsonian Center for Astrophysics, the Las Cumbres Observatory Global Telescope Network Incorporated, the National Central University of Taiwan, the Space Telescope Science Institute, the National Aeronautics and Space Administration under Grant No. NNX08AR22G issued through the Planetary Science Division of the NASA Science Mission Directorate, the National Science Foundation Grant No. AST-1238877, the University of Maryland, Eotvos Lorand University (ELTE), the Los Alamos National Laboratory, and the Gordon and Betty Moore Foundation.
\end{acknowledgements}

\bibliographystyle{aa}
\bibliography{my_bib} 

\newpage

\begin{appendix}
\section{Timing residuals}
\label{app:tr}
\begin{figure}[h]
    \centering
    \includegraphics[width=0.9\linewidth]{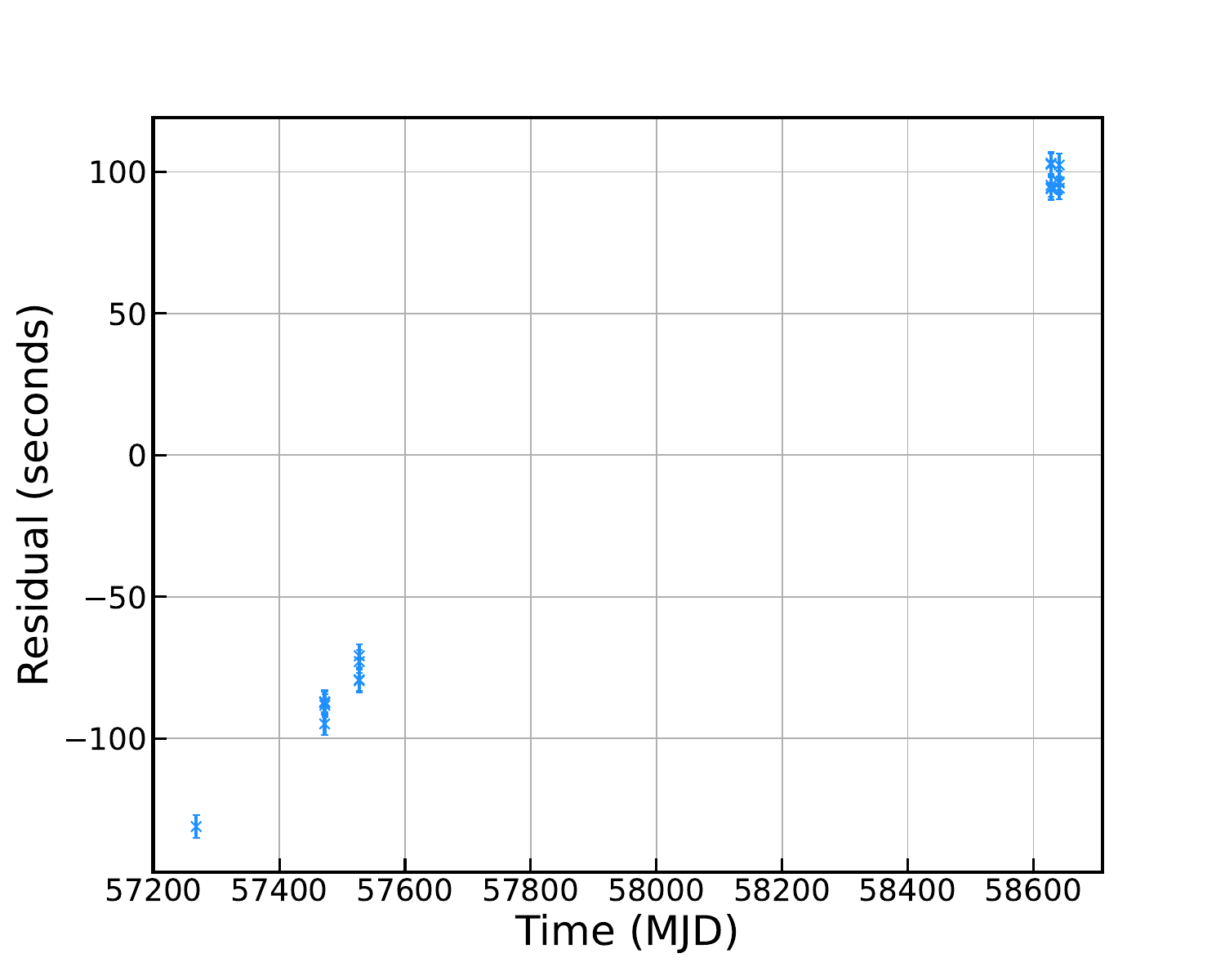}
    \caption{Residuals for the first timing solution in Tab.\,\ref{tab:timing-models}.}
    \label{fig:t1}
\end{figure}

\begin{figure}[h]
    \centering
    \includegraphics[width=0.9\linewidth]{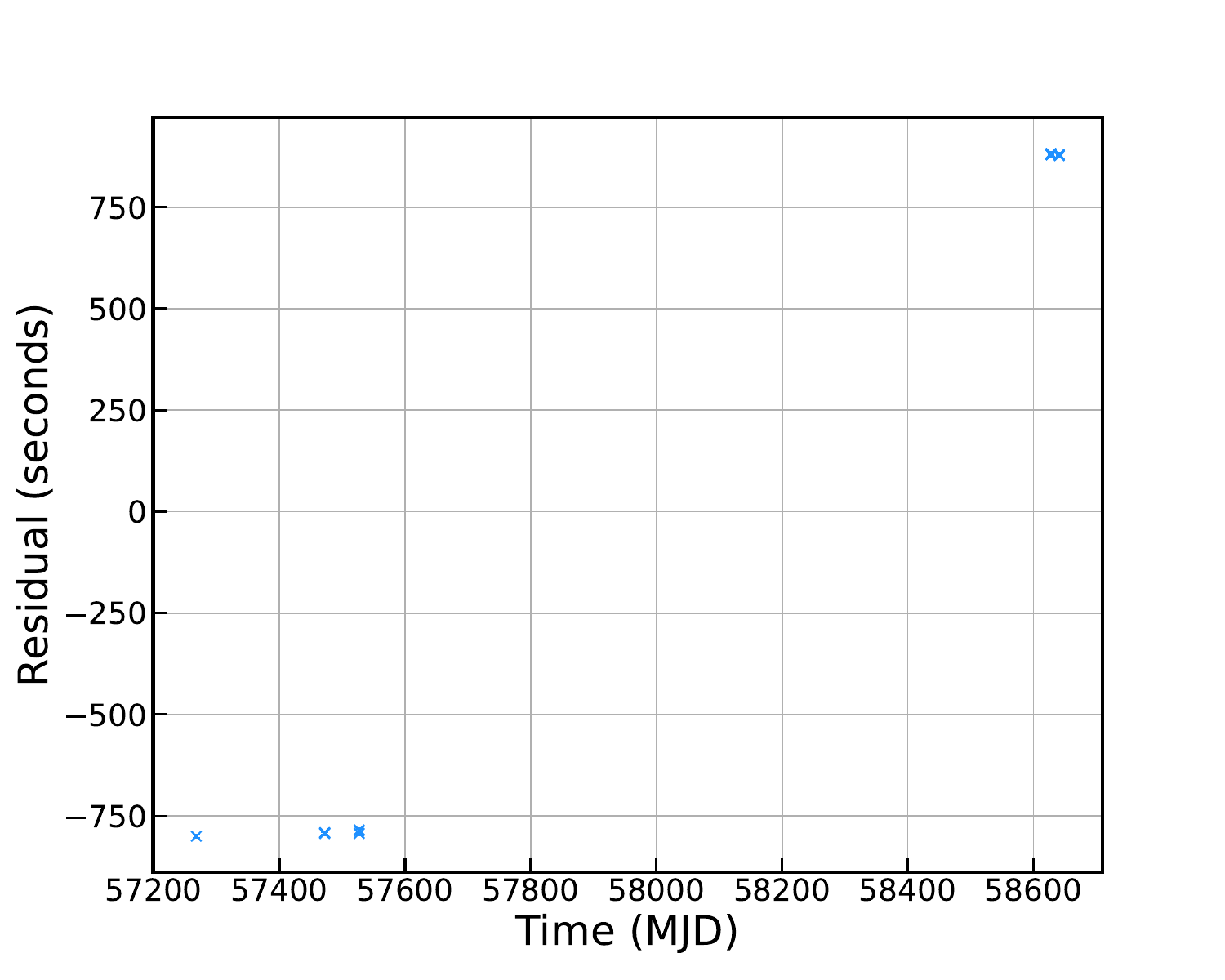}
    \caption{Residuals for the second timing solution in Tab.\,\ref{tab:timing-models}.}
    \label{fig:t2}
\end{figure}

\begin{figure}[h]
    \centering
    \includegraphics[width=0.9\linewidth]{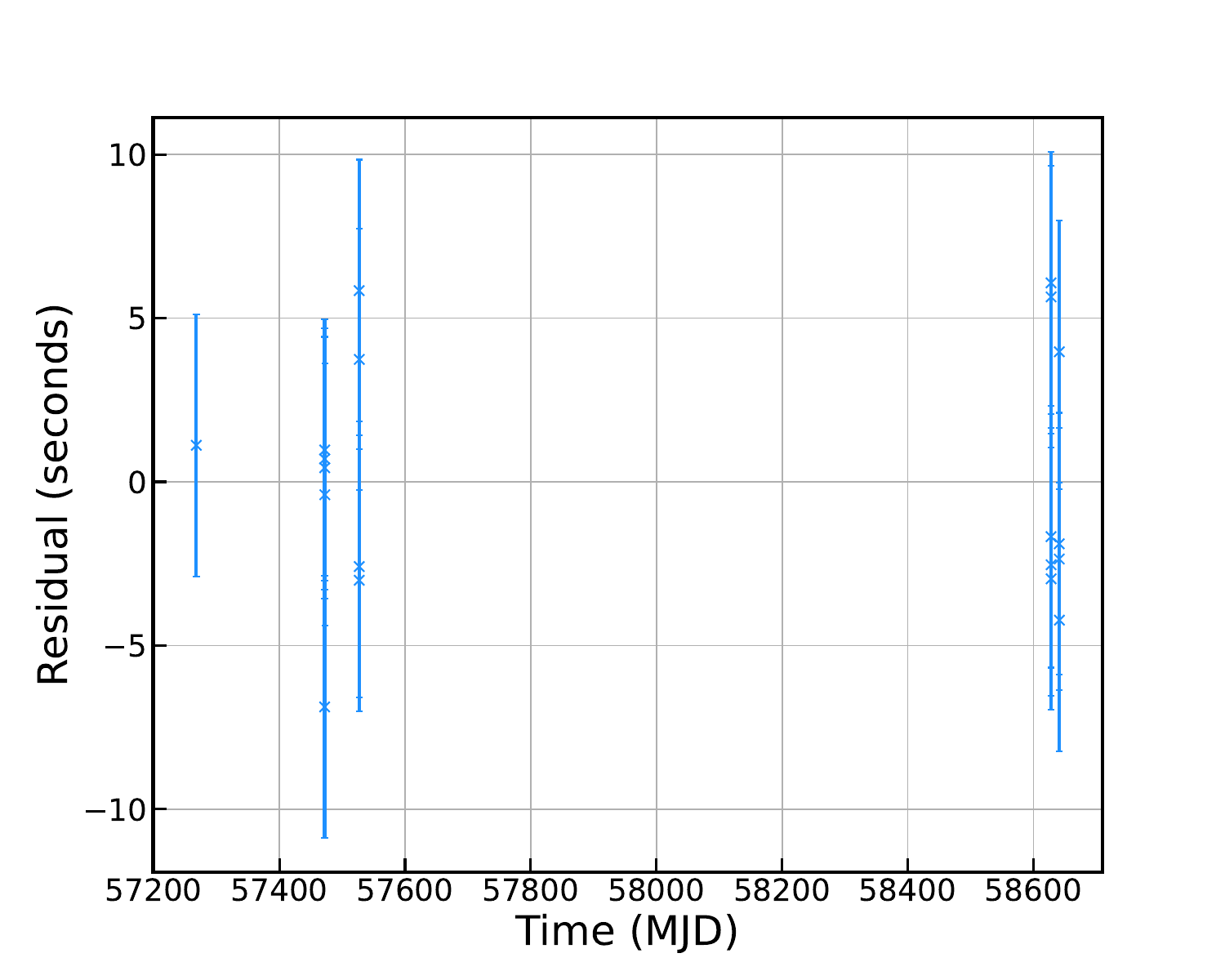}
    \caption{Residuals for the third timing solution in Tab.\,\ref{tab:timing-models}.}
    \label{fig:t3}
\end{figure}

\begin{figure}[h]
    \centering
    \includegraphics[width=0.9\linewidth]{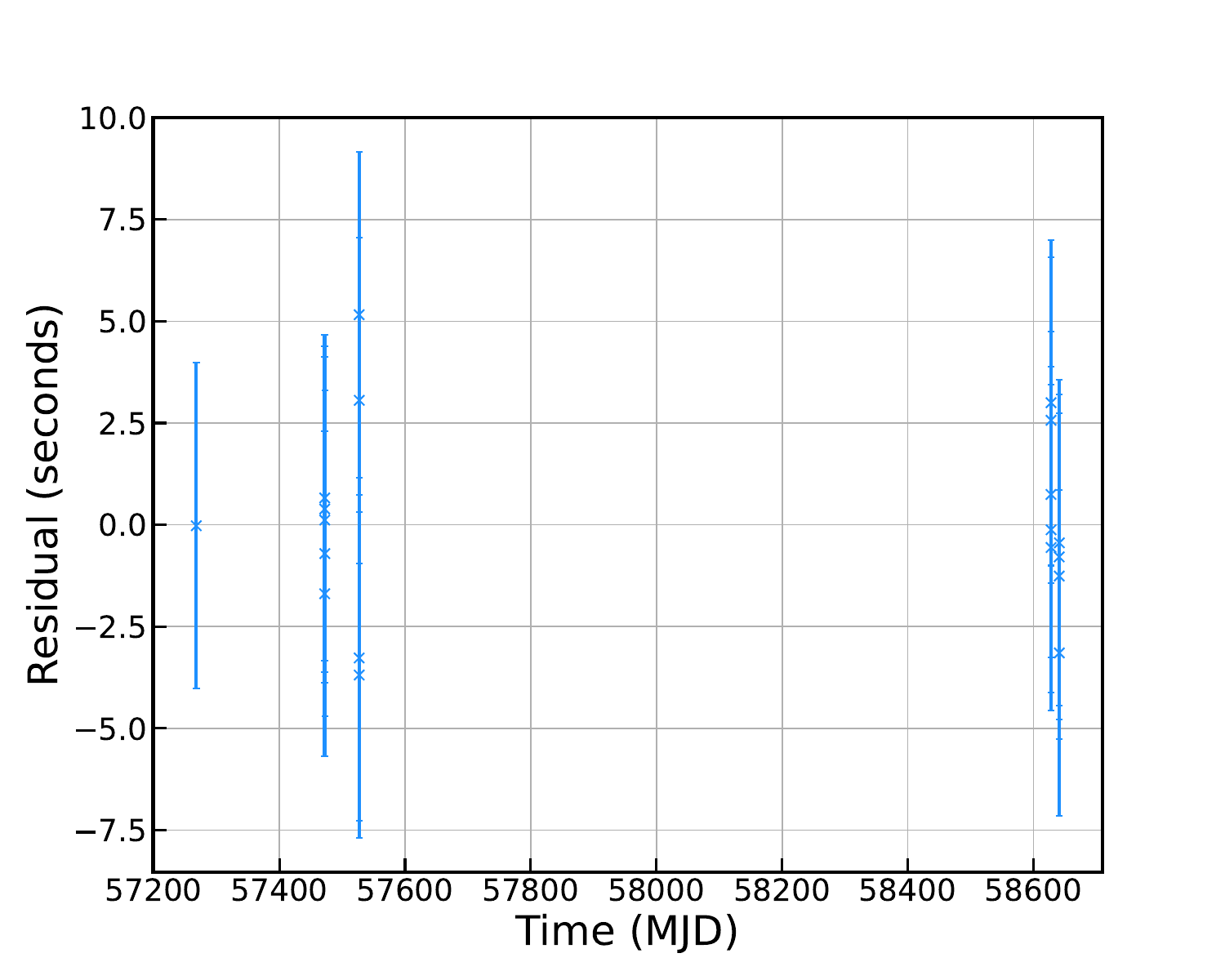}
    \caption{Residuals for the fourth timing solution in Tab.\,\ref{tab:timing-models}.}
    \label{fig:t4}
\end{figure}

This appendix contains the residuals for the four different timing solutions presented in Tab.\,\ref{tab:timing-models}. Fig.\,\ref{fig:t1} shows the residuals for a timing model with a period of 841\,s. Fig.\,\ref{fig:t2} shows the residuals for a timing model with the longer period of 4206\,s and an interpulse phase offset of 0.6. Fig.\,\ref{fig:t3} shows the residuals for a timing model with an 841\,s period and a period derivative of $-9\times10^{-12}$\,s\,s$^{-1}$. Fig.\,\ref{fig:t4} shows the residuals for a timing model with a period of 4206\,s, an interpulse phase offset of 0.6, and a phase jump between 2015--2016 and 2019 of 0.614.

\end{appendix}
\end{document}